\documentclass[sigconf,10pt]{acmart}
\renewcommand\footnotetextcopyrightpermission[1]{}
\acmConference{Published to Arxiv}{Copyright Oracle}{2021}

\usepackage{xspace}

\newcommand{\etal}{{\it et al.}\xspace}

\setcopyright{rightsretained}

\begin{document}

\title[Modeling memory bandwidth patterns with performance counters]{Modeling memory bandwidth patterns on NUMA machines with performance counters}

\author{Daniel Goodman}
\affiliation{%
  \institution{Oracle Labs}
}
\email{daniel.goodman@oracle.com}
\author{Roni Haecki}
\authornote{Work carried out while at Oracle Labs}
\affiliation{%
  \institution{Oracle Labs}
}
\author{Tim Harris}
\authornote{Work carried out while at Oracle Labs.}
\affiliation{%
  \institution{Oracle Labs}
}

\begin{abstract}
Computers used for data analytics are often NUMA systems with multiple sockets per machine, multiple cores per socket, and multiple thread contexts per core. To get the peak performance out of these machines requires the correct number of threads to be placed in the correct positions on the machine. One particularly interesting element of the placement of memory and threads is the way it effects the movement of data around the machine, and the increased latency this can introduce to reads and writes. In this paper we describe work on modeling the bandwidth requirements of an application on a NUMA compute node based on the placement of threads. The model is parameterized by sampling performance counters during 2 application runs with carefully chosen thread placements. Evaluating the model with thousands of measurements shows a median difference from predictions of  2.34\% of the bandwidth. The results of this modeling can be used in a number of ways varying from: Performance debugging during development where the programmer can be alerted to potentially problematic memory access patterns; To systems such as Pandia which take an application and predict the performance and system load of a proposed thread count and placement; To libraries of data structures such as Parallel Collections and Smart Arrays that can abstract from the user memory placement and thread placement issues when parallelizing code.
\end{abstract}

\settopmatter{printacmref=false}
\maketitle

\section{Introduction}
\begin{figure}
\begin{center}
\includegraphics[trim=0 700 0 0, clip,width=\linewidth,page=20]{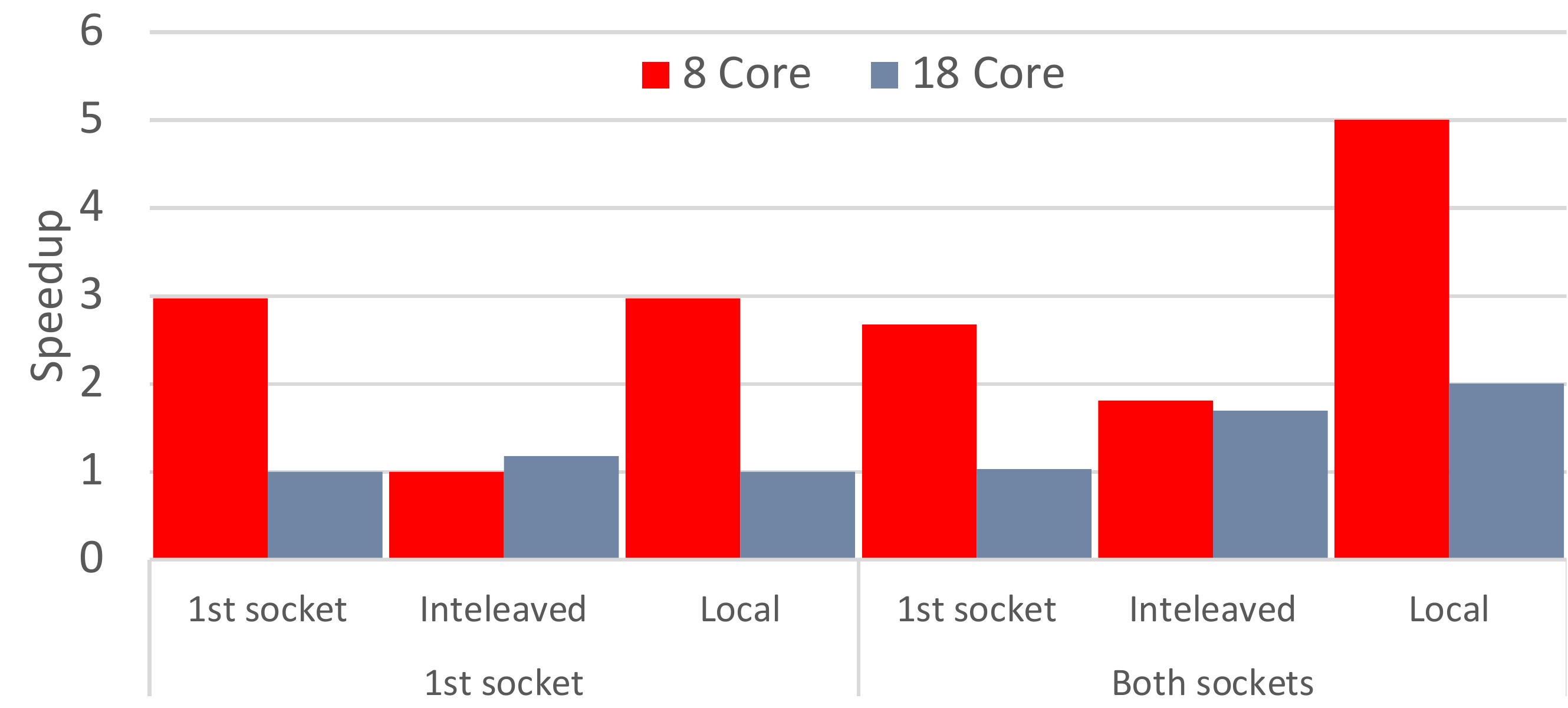}
\end{center}
\caption{Graph showing the performance of a memory intensive application on different dual socket Intel Xeon machines with different thread and memory placements.  Speedup is relative to the slowest configuration for each machine. Results are labeled by the memory placement, 1st socket, interleaved, or local, and then by the thread placement, 1 socket, or both sockets.}
\label{fig:motivation}
\end{figure}


Modern NUMA computers with multiple sockets per machine, multiple cores per socket, and multiple thread contexts per core are often used for data analytics type workloads. To get the peak performance out of these machines requires the correct number of threads to be placed in the correct positions on the machine. One particularly interesting element of the placement of memory and threads is the way it effects the movement of data around the machine. This effects the latency of reads and writes both through the inherent different latencies of different connections, but also through the different bandwidths of the connections and the effect if these become saturated. This paper describes work on modeling the bandwidth requirements of an application on a NUMA compute node based on the placement of threads. The model is parameterized by sampling performance counters during 2 application runs with carefully chosen thread placements. Previous work using performance counters has tended to focus on the IPC and LLC miss counters, but as the number of performance counters available on modern processors increases~\cite{Zellweger2016} the opportunity to construct more intricate models also increases.  Evaluating the model over many thousands of measurements has shown a median difference between the predicted and the measured bandwidth of 2.34\% of the bandwidth, with larger errors primarily restricted to applications that transfer little data.  The results of the modeling can be used in a number of ways varying from: Performance debugging during development where the programmer can be alerted to potentially problematic memory access patterns; To systems such as Pandia~\cite{Goodman:2017:pandia} which take an application and predict the performance and system load of a proposed thread count and placement; To libraries of data structures such as Parallel Collections~\cite{scalaParallelCollections} and Smart Arrays~\cite{smartArrays} that can abstract from the user memory placement and thread placement issues when parallelizing code.

\begin{figure}
\begin{center}
\includegraphics[trim=0 712 0 0,clip,width=\linewidth,page=16]{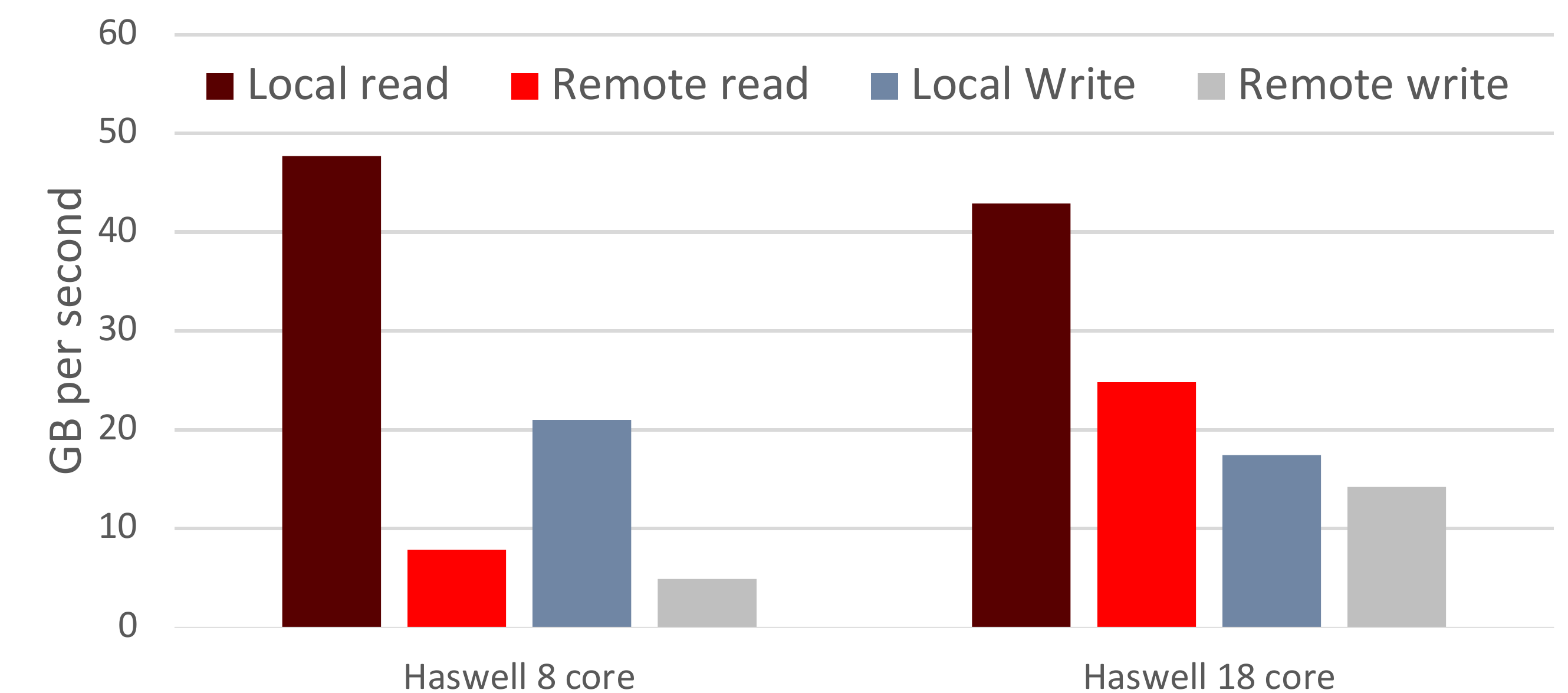}
\end{center}
\caption{Graph of the different memory bandwidths available on the test systems.}
\label{fig:commuunicationSpeed}
\end{figure}

As motivation for this work consider the graph in Figure~\ref{fig:motivation}. This shows the performance of different thread and memory placements on 2 different 2 socket Intel Haswell machines. Each run is normalized relative to the slowest run on a given machine. The same memory intensive synthetic benchmark is used for all runs and was run with enough threads to place a single thread on each core of a single socket, 8 and 18 threads respectively. When running with 2 sockets these threads are split evenly with each thread still placed in its own core. The memory placements are: 1) all memory on the first socket; 2) Memory interleaved between sockets at the granularity of a page giving 50\%  remote accesses; 3) All of a threads memory located locally to the thread giving 0\% remote accesses. From this we can see that when using a single socket for the 18 core system there is little difference between accessing data remotely and accessing it locally with the CPU acting as the limiting factor, while for the 8 core system there is a 3x slowdown relative to the quickest placement. We can also observe that when running an application with shared memory, the fastest placement for the 18 core machine is to spread the threads and the data evenly across the machine interleaving the memory to spread the bandwidth load evenly. For the 8 core machine peak performance is achieved by keeping all the data and threads on a single socket avoiding remote communication. It is clear from this that the 18 core machine is far more forgiving of thread and memory placement, however as shown in Figure~\ref{fig:commuunicationSpeed} the 8 core machine has a higher bandwidth to the local memory, and is a substantially cheaper machine with a suggested retail price per CPU of \$667 vs \$4115. This means that if the placement of memory and threads can be correctly organized there is the potential to save both time and money on memory limited applications. Figure~\ref{fig:commuunicationSpeed} also demonstrates the dramatic difference in bandwidth for 2 apparently similar machines. 

Having demonstrated the effects that different memory placements can have on a system we will now consider how this model could be used in the use cases mentioned earlier. 

\emph{Performance prediction}
Systems that seek to model the performance of a workload in a given configuration need to be able to predict the resource demands. Currently in systems such as Pandia they either use a static placement for all applications, or measure the distribution of bandwidth during one of the measurement runs and assume that it will stay the same throughout all further thread placements. The use of this model would allow for more detailed bandwidth requirements to be added so allowing for a more accurate prediction of the performance with each possible thread count and placement.

\paragraph{Libraries of data structures}
When libraries such as Smart Arrays abstract the memory placement from the user, the library needs to make decisions about how best to layout the memory. Currently they make the assumption that when the collection requires bandwidth for data processing the application will not be using bandwidth for anything else. However the assumption that all the data used by the application at the time will be in collections belonging to the library is a restrictive one. The use of a bandwidth model would provide the opportunity to model the bandwidth requirements and make allowances for this at run time when placing the data stored by the library into memory.

\paragraph{Debugging and development}
Applications will typically run in many environments and performance critical applications have to be tested in each environment that they wish to run in. By modeling the bandwidth requirements of the application with different thread placements and against different hardware descriptions, it would be possible to flag up potential problems to the programmer before the application reaches the testing stage, so allowing an earlier fix.

\vspace {0.4cm}
 Our contributions are the bandwidth model, the techniques for parameterizing it to fit real programs, and the evaluation of the model. 

The rest of this paper is structured as follows: We will first introduce an outline of the system that we are modeling; Then we will look at the way we model the bandwidth; Having introduced the model we will look at how it can be used to predict the bandwidth requirements of a thread placement; Before we consider how we measure the values for the model for a given application; We then look at the stability and accuracy of this modeling across different hardware with a range of synthetic and real benchmarks; Before concluding with thoughts on future work and looking at related work.

\section{Target Hardware}
This work is aimed at multi-socket NUMA machines with performance counters to record their activity. It was carried out on Intel Xeon machines. An idealized example of a 2 socket machine can be seen in Figure~\ref{fig:baseMachine} with a memory bank attached to each socket via memory channels, and the sockets connected via an interconnect. Information about the performance of applications is captured via performance counters the processors provide through tools such as Performance Counter Monitor (PCM). While some variation appears between manufactures, the counters offered by manufactures are sufficiently similar that we believe that the techniques described here can be applied to other hardware with minimal modifications. 

\subsection{Performance counters}
\begin{figure}
\begin{center}
\includegraphics[trim=0 619 0 0, clip,width=\linewidth,page=2]{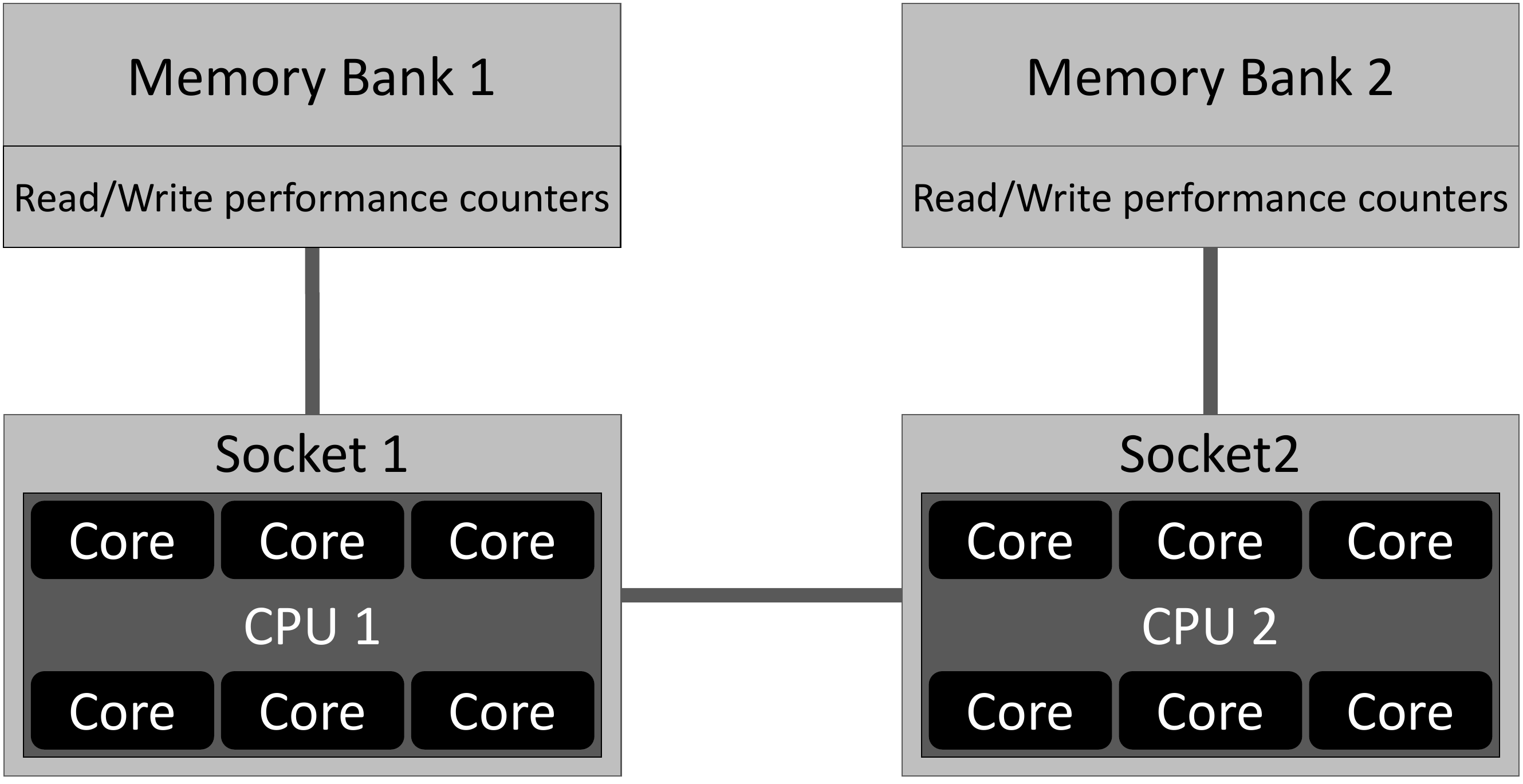}
\end{center}
\caption{Example of a 2 socket NUMA machine with banks of memory attached to each socket, and each socket containing a CPU with 6 cores. We will use the same machine in all our examples, but for visual clarity we will dispense with drawing the cores. Each memory bank is connected to its respective socket by a memory channel, and an interconnect joins the sockets. While the performance counters recording the data transfer are almost certainly located on the memory controller contained within the CPU, we have drawn them on the memory banks here to emphasize that on the Intel processors used for this work the counters report from the perspective of the memory bank, not the processor regarding local and remote accesses.}
\label{fig:baseMachine}
\end{figure}

The counters of interest in this work cover the time elapsed, the number of instructions executed, and the volume of data read or written to each memory bank separated out into data to and from the local socket and data to and from remote sockets. While the performance counters recording the volume of data to and from the memory banks are almost certainly located on the memory controller contained within the CPU, in Figure~\ref{fig:baseMachine} we have drawn them on the memory banks to emphasize that on the Intel processors used for this work the counters report from the perspective of the memory bank, not the processor regarding local and remote accesses. So for example if we have 2 threads on CPU1 and 1 thread on CPU 2 all running at the same speed, and all sending $\frac{1}{2}$ of their accesses to each of the memory banks, then from the point of view of the CPUs $\frac{1}{2}$ of memory accesses are to remote locations, and $\frac{1}{2}$ are to local locations. However from the point of view of memory banks 1 and 2, $\frac{2}{3}$ and $\frac{1}{3}$ of the accesses are local respectively with the remainder being remote accesses, and it is this view that is reported by the program counters. 

\subsubsection{lessons learned}
When selecting the performance counters we noted two sets of counters that we initially thought would be useful but it turned out were insufficient for our use case:

\paragraph{Quick Path Interconnect (QPI)} The first of these was the QPI counters. We had hoped that these would allow us to observe directly how much of our data was moving along which interconnects. However, there is a substantial amount of additional traffic that makes use of the QPI, much of which appears to die away when the QPI is required for moving application data. So this traffic is not a limiting factor on the performance of an application, but does make for a very noisy signal when trying to model the bandwidth. For this reason we instead chose to look at the traffic to the memory banks which is considerably less noisy as a metric.

\paragraph{Instructions Per Cycle (IPC)} Many performance counter libraries still include IPC as one of the performance counters that users can request, and while the value returned is the IPC, as the frequency of the chips can change at a very fine granularity this information is extremely misleading without a corresponding record of the chip frequency. To over come this we instead opt to record the number of instructions executed and the time in which they were executed.

\section{Bandwidth model}
Our model of the bandwidth utilization across a machine is constructed by splitting the usage into 4 classes of memory access pattern that can be combined to describe the memory access pattern of the application threads. The fraction of each threads memory accesses that can be attributed to each of these patterns is measured through 2 profiling runs and we will call this a bandwidth signature. The resulting signature can then be used to apply bandwidth requirements to any thread placement. Separate signatures are constructed for reads and writes, but the measurements required for these two signatures are taken during a single set of runs. The access types are described here on a system with $s$ sockets and executing a workload using $n$ threads:

\begin{description}
\item[Static] Access to memory that is allocated on the RAM attached to a single socket but used by all threads. For example if the master thread loads the input data or the output array is allocated on a single socket.
\item[Local] Access to memory that is only accessed by the threads on the same socket as the memory. For example a replicated data structures, or thread local data.
\item[Interleaved] Access to memory that is allocated evenly across the used sockets such that each socket has $\frac{1}{s}$ of the data. For example if the interleave flag has been used in \texttt{numactl}.
\item[Per-thread] Access to memory where each thread allocates $\frac{1}{n}$ of the memory locally, but the memory is used by all threads. For example if each thread loads $\frac{1}{n}$ of the data, or the threads are are constructing and using a shared data structure such as a tree.
\end{description}

\begin{figure}
\begin{center}
\includegraphics[trim=0 619 0 0, clip,width=\linewidth,page=3]{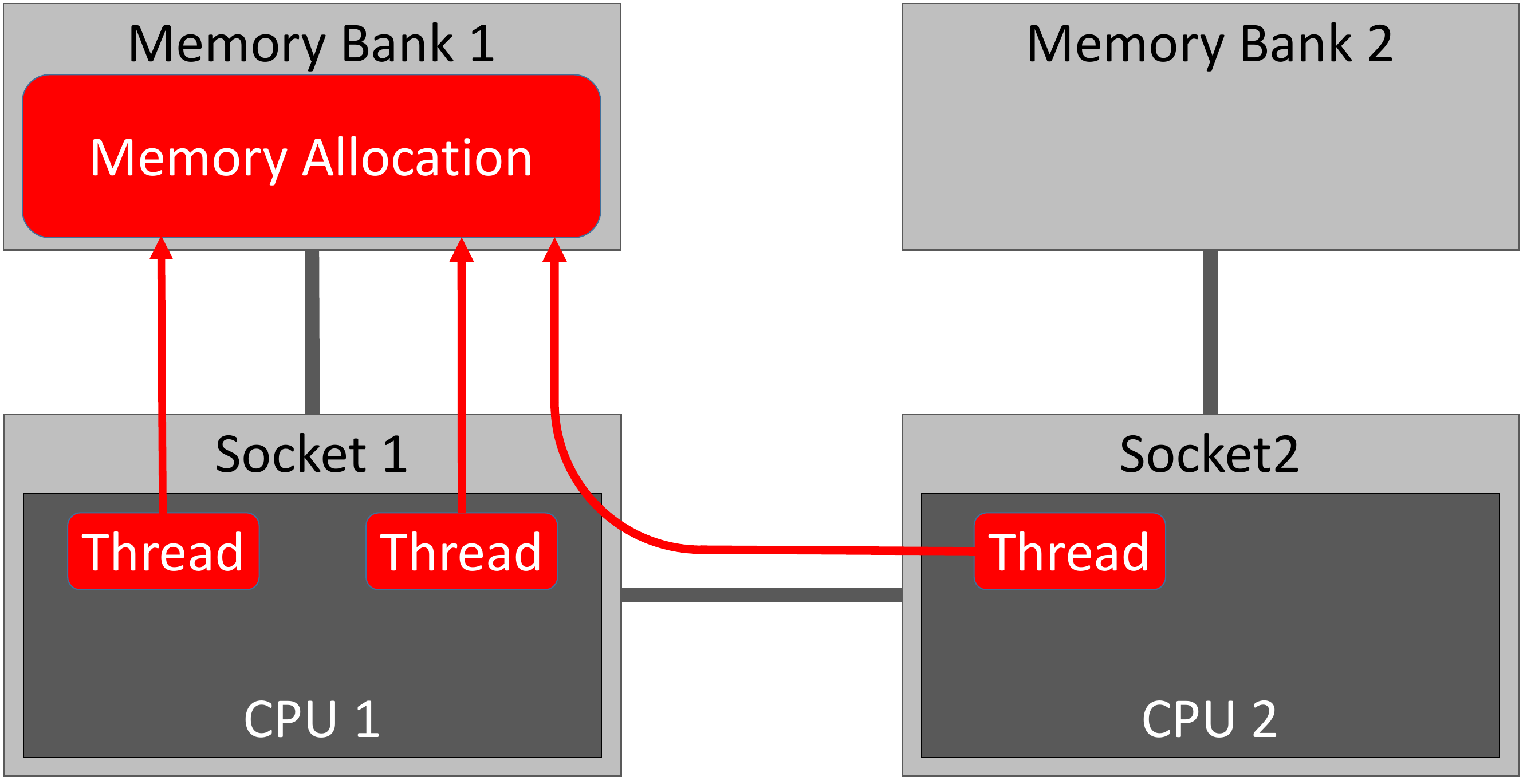}
a) Static
\includegraphics[trim=0 619 0 0,clip,width=\linewidth,page=4]{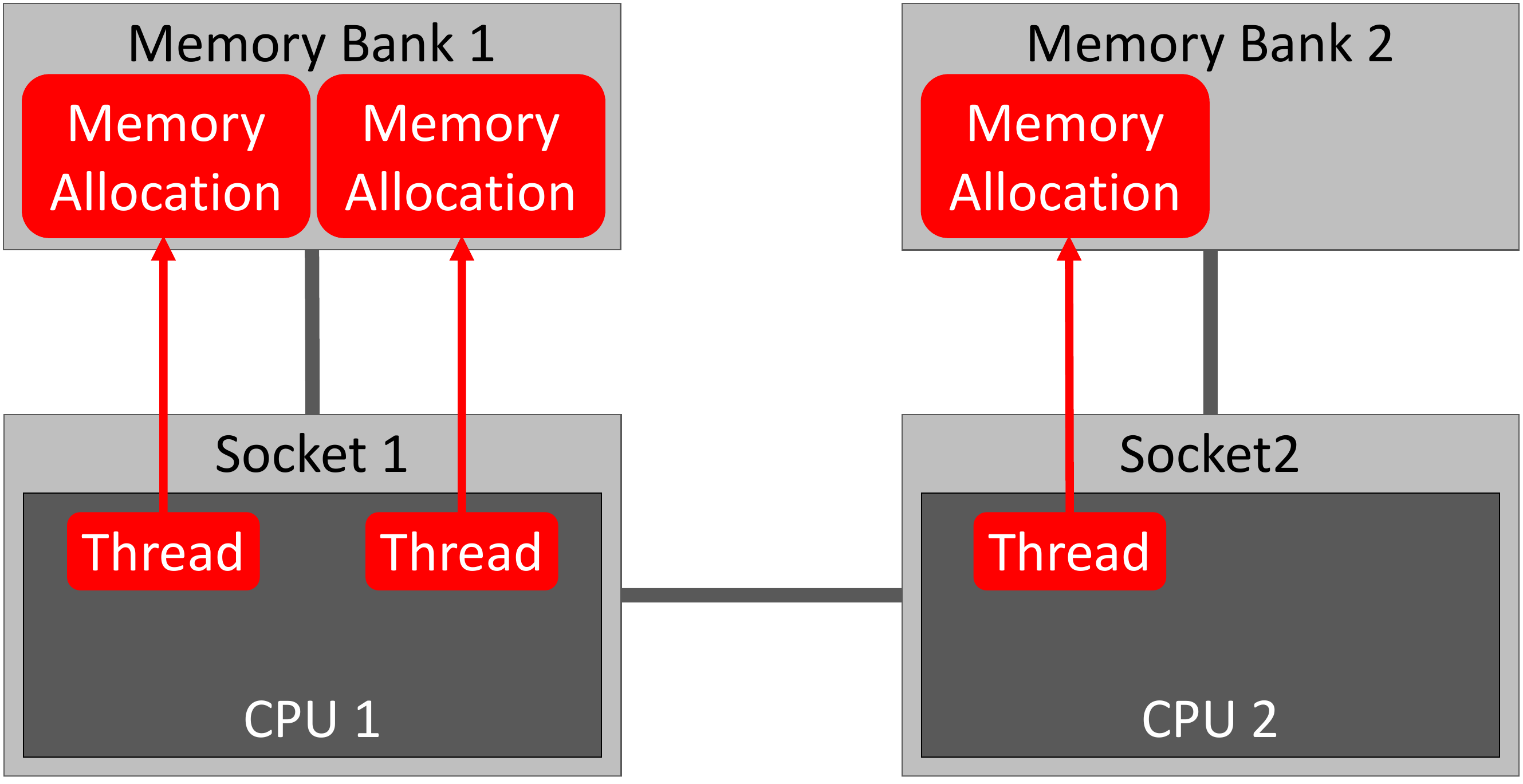}
b) Local
\includegraphics[trim=0 619 0 0,clip,width=\linewidth,page=5]{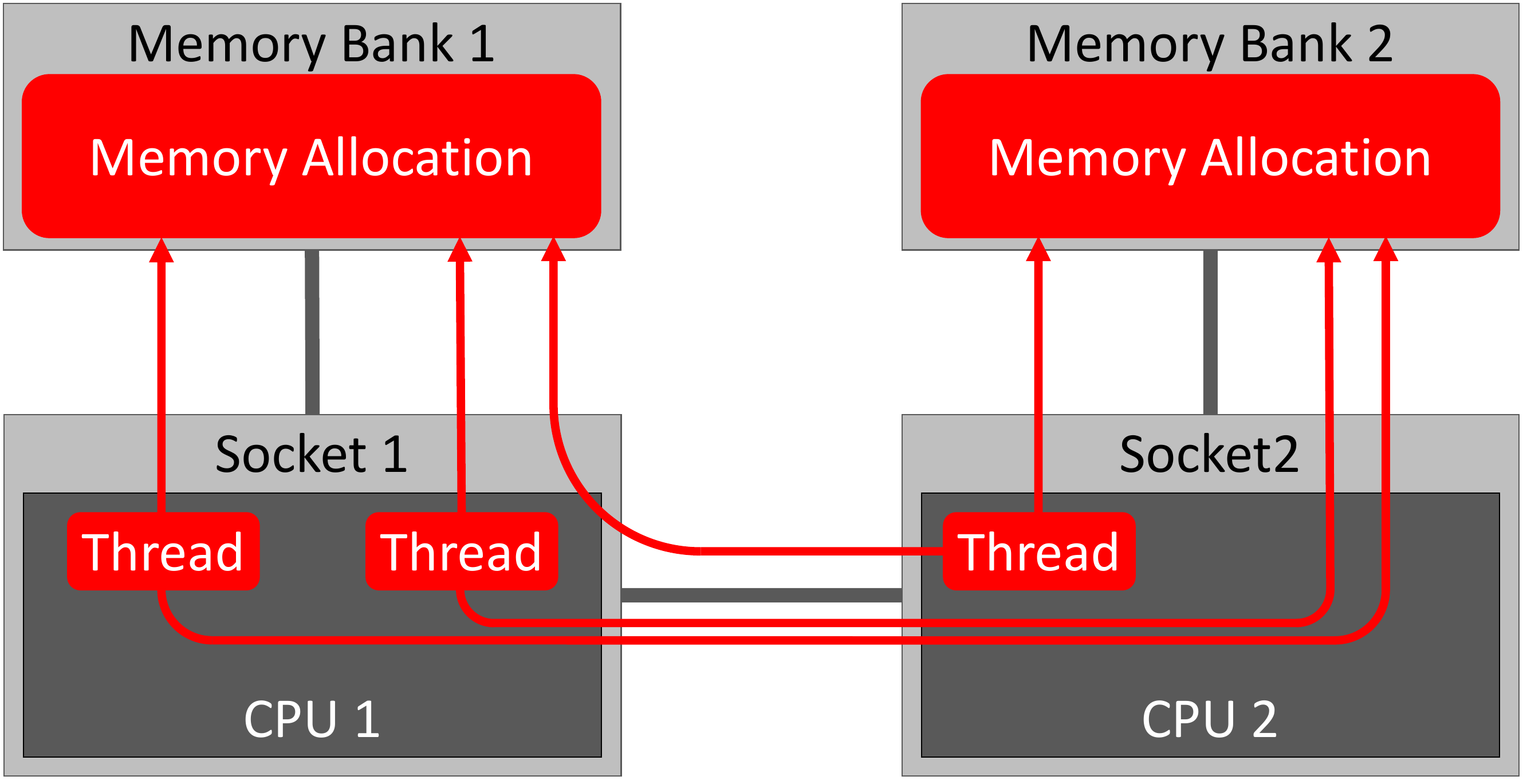}
c) Interleaved
\includegraphics[trim=0 619 0 0,clip,width=\linewidth,page=6]{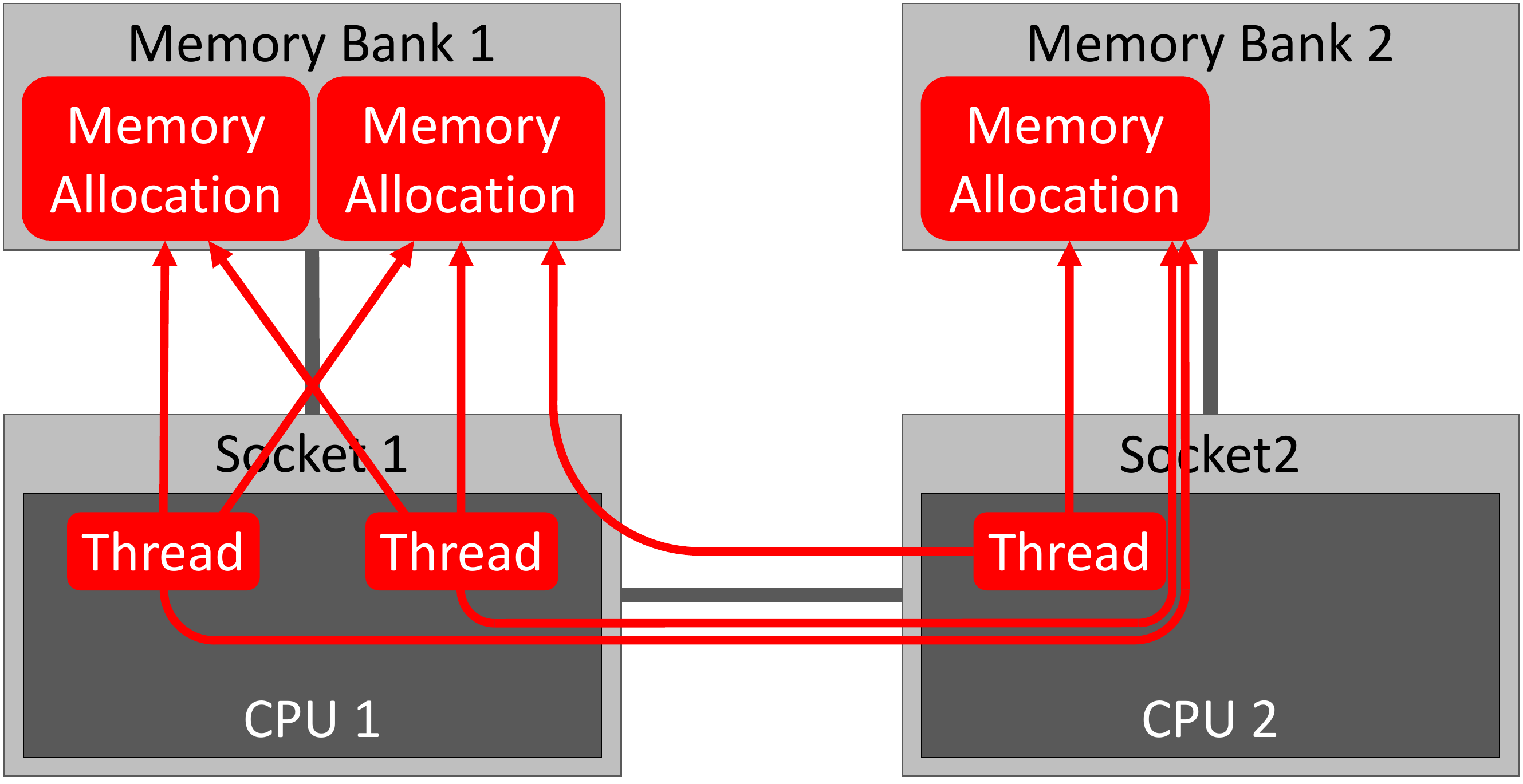}
d) Per thread
\end{center}
\caption{Examples of the 4 different types of memory placement types that we model. In this case they are an application with 3 threads placed on a machine with 2 sockets.}
\label{fig:accessPatterns}
\end{figure}

\noindent Examples of these patterns can be seen in Figure~\ref{fig:accessPatterns}.

We view that the read bandwidth and the write bandwidth requirements for each workloads is made up of a mix of these 4 classes of access pattern. To encode this for both reads and writes we use 3 properties in the range $[0\ldots1]$ describing the fraction of the accesses that is Per thread, Local, and Static. We call these the \emph{Per thread fraction}, \emph{Local fraction}, and \emph{Static fraction}. Any remaining bandwidth is deemed to be Interleaved. The sum of the three fractions must be $\leq 1$. This gives 6 properties in total, 3 for reads and 3 for writes. These 6 are augmented by a property for read bandwidth and a property for write bandwidth that records which socket the static bandwidth is associated with. We call this the \emph{Static socket}. Together these properties make up the bandwidth signature of the application. 

Before considering how to calculated the 4 properties that make up the model for reads and the 4 properties for writes we now consider how to apply them to construct the bandwidth requirements for different thread placements.

\section{Applying bandwidth signature to a thread placement}
As discussed the model describes how the bandwidth from a thread on a given socket should be distributed across the system, the total volume of data for each thread will need to be calculated independently. For example in Pandia this is achieved by taking the bandwidth requirement of a single thread, applying this to every thread and then scaling the bandwidths on a thread by thread basis to allow for changes in thread performance due to issues such as resource saturation. 

To demonstrate the applying the signature we will use an example which has the properties Static~socket~= 2, Static~fraction~=~0.2, Local~fraction~=~0.35, and Per~thread~fraction~=~0.3 with  $1-(0.2 + 0.35+0.3)$ giving 0.15 as the value of the fraction of the bandwidth that is due to interleaved placements.  We will consider a placement of 4 threads on a two socket machine with 3 threads placed on the first socket and 1 thread on the second socket. 

One way to think about this is as a matrix computation where we have a matrix for each type of memory traffic. Within these matrices  each row represents  data traveling to or from a CPU, and each column represents data traveling to or from a memory bank. Each cell represent the fraction of the data traveling to or from a given CPU and a given memory bank. As a result the sum of each row will be 1. The four matrices that are then constructed as follows:
\begin{description}
 \item[Static] The matrix for static memory accesses represents all traffic going to a single memory bank and so consists of the column identified by the static socket property containing 1's and all other columns containing zeros.
 \item[Local] The matrix for local memory accesses models all data accesses from a socket going to their corresponding memory bank. This is represented by an identity matrix.
 \item[Per Thread] The matrix for per thread data consists of a series of columns weighted by the fraction of the threads that are on each socket, so the weights for column $i$ can be calculated by $\frac{n_i}{\sum_{j=1}^{s}n_j}$ where $n_i$ is the number of threads on socket $i$.
 \item[Interleaved] The matrix for interleaved data models all accesses being spread evenly across the system. There for cells where both the memory bank and the CPU are from used sockets contain $\frac{1}{s}$ and the other cells contain 0 where $s$ is the number of sockets in use in the placement.
 \end{description}

These matrices can the be scaled by their respective fractions, and summed. This results in a single matrix mapping a threads socket to the fraction of its bandwidth it is predicted place on each link to the memory banks. Figure~\ref{fig:calculate} shows this process for the worked example. 

\begin{figure}
\begin{center}
\includegraphics[trim=0 3410 550 0,clip,width=\linewidth,page=1]{figures/figures.pdf}
\end{center}
\caption{Diagram demonstrating for our example how calculation of the fractions of bandwidth to different memory banks can be treated as a matrix computation. This example is for a 2 socket machine with 3 threads on the first socket and 1 thread on the second. Note that every row sums to 1, but not every column.}
\label{fig:calculate}
\end{figure}

\section{Measuring an applications bandwidth signature}
Having looked at how the model can be used we will now consider how to measure the parameters required for the model. As our aim is to model the bandwidth utilization of the various elements of the machine it is not sufficient to just observe where the memory is allocated as a relatively small piece of memory may be accessed frequently while a larger piece of memory may be seldom accessed. It is also not sufficient to just measure the bandwidth requirements of a single thread in a single location and assume that the pattern of accesses it displays will not change as the threads position changes.

\begin{figure}
\begin{center}
\includegraphics[trim=0 829 637 0,clip,width=\linewidth,page=22]{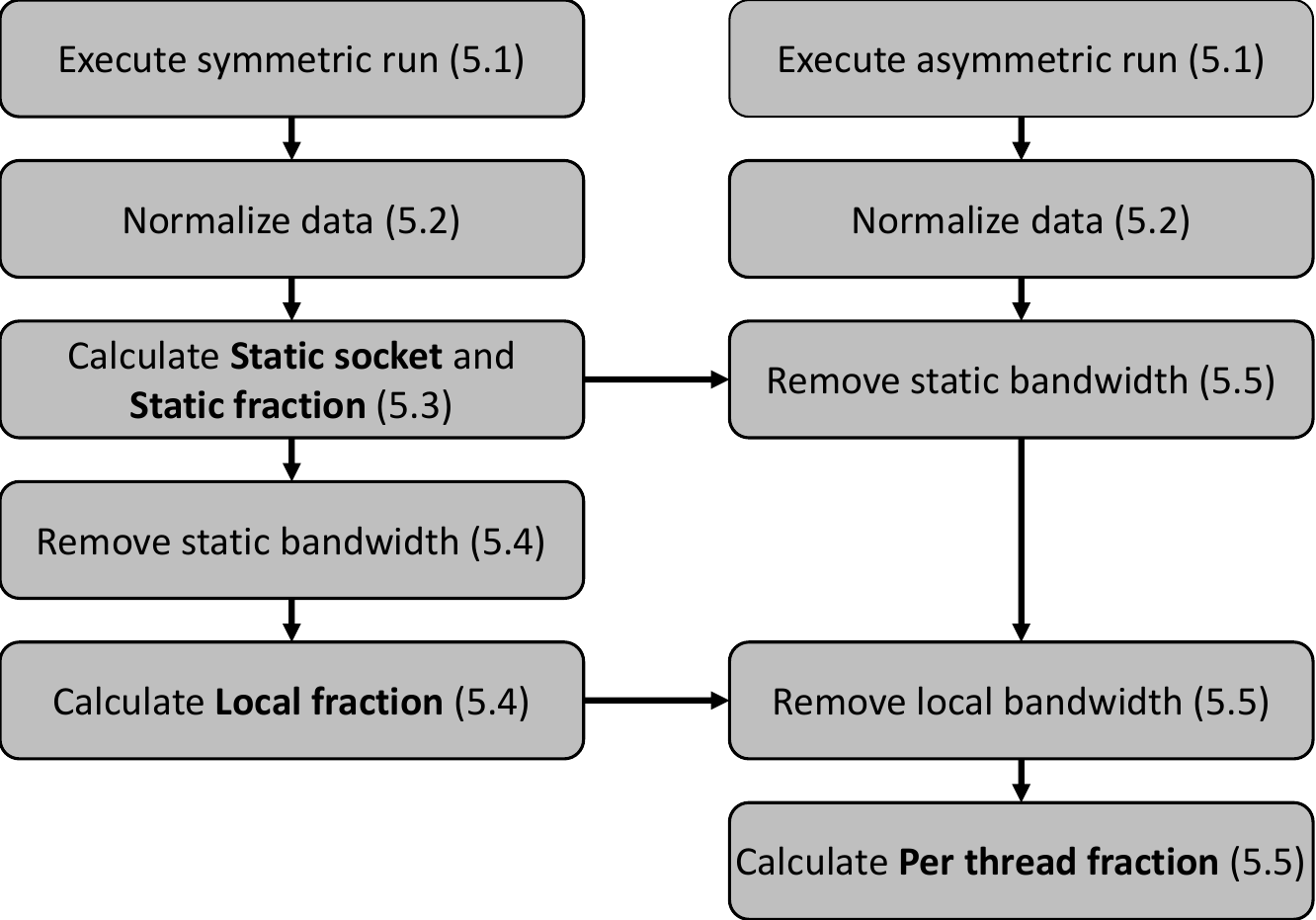}\\
\end{center}
\caption{Flow diagram of the steps involved in constructing a bandwidth signature.}
\label{fig:flow-diagram}
\end{figure}

The techniques to calculate bandwidth signature can be applied to differing numbers of sockets, but for clarity here we will describe how to calculate the properties using just 2 sockets. A flow diagram of the steps involved in calculating the signature can be seen in Figure~\ref{fig:flow-diagram}.

\subsection{Profiling runs}
To calculate the signature we take advantage of program counters that for the memory attached to each socket report the volume of data sent to and from the local CPU and the volume of data sent to and from the remote CPUs. Using these program counters we collect data from 2 runs of the workload. 

\begin{figure}
\begin{center}
\includegraphics[trim=0 700 0 0,clip,width=\linewidth,page=7]{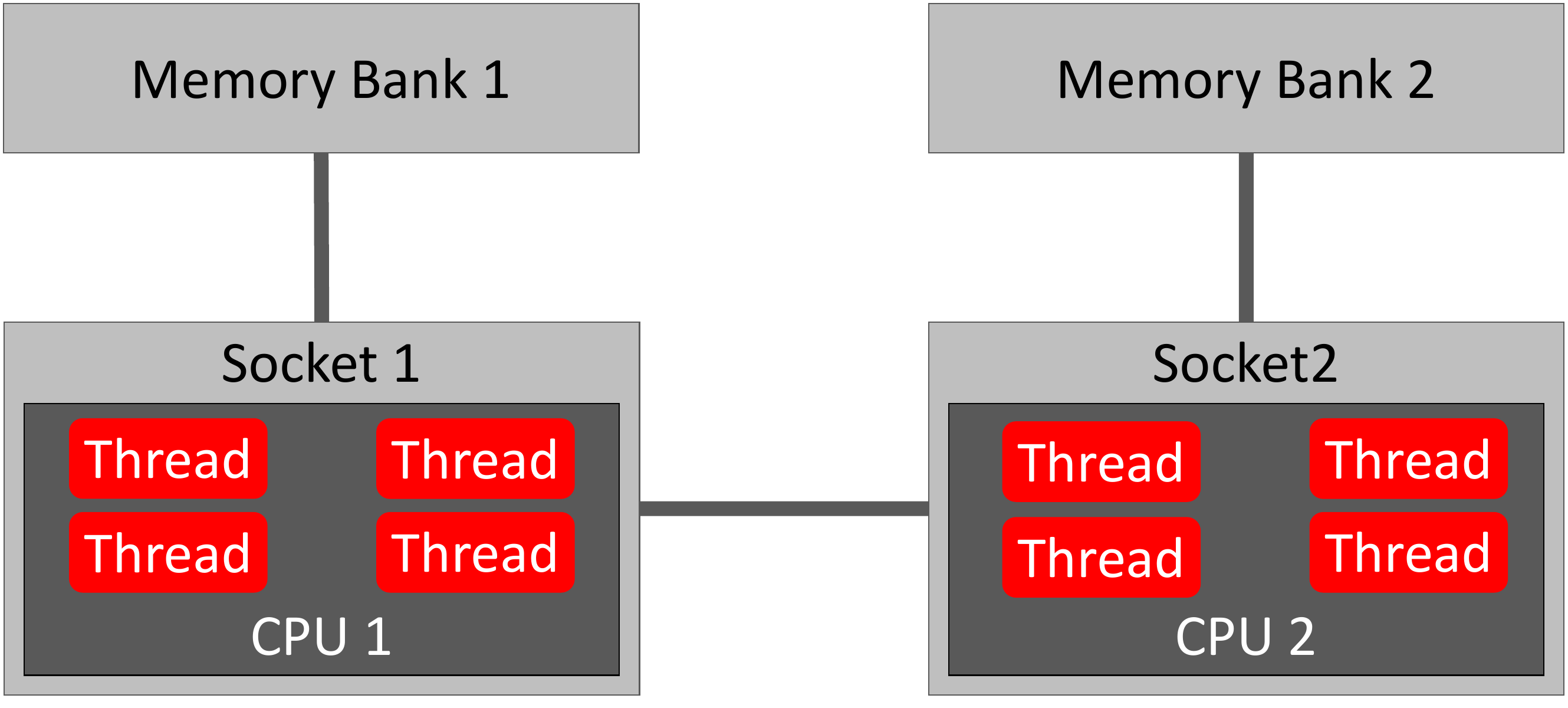}\\
a) Symmetric\\
\vspace{0.2cm}
\includegraphics[trim=0 700 0 0,clip,width=\linewidth,page=8]{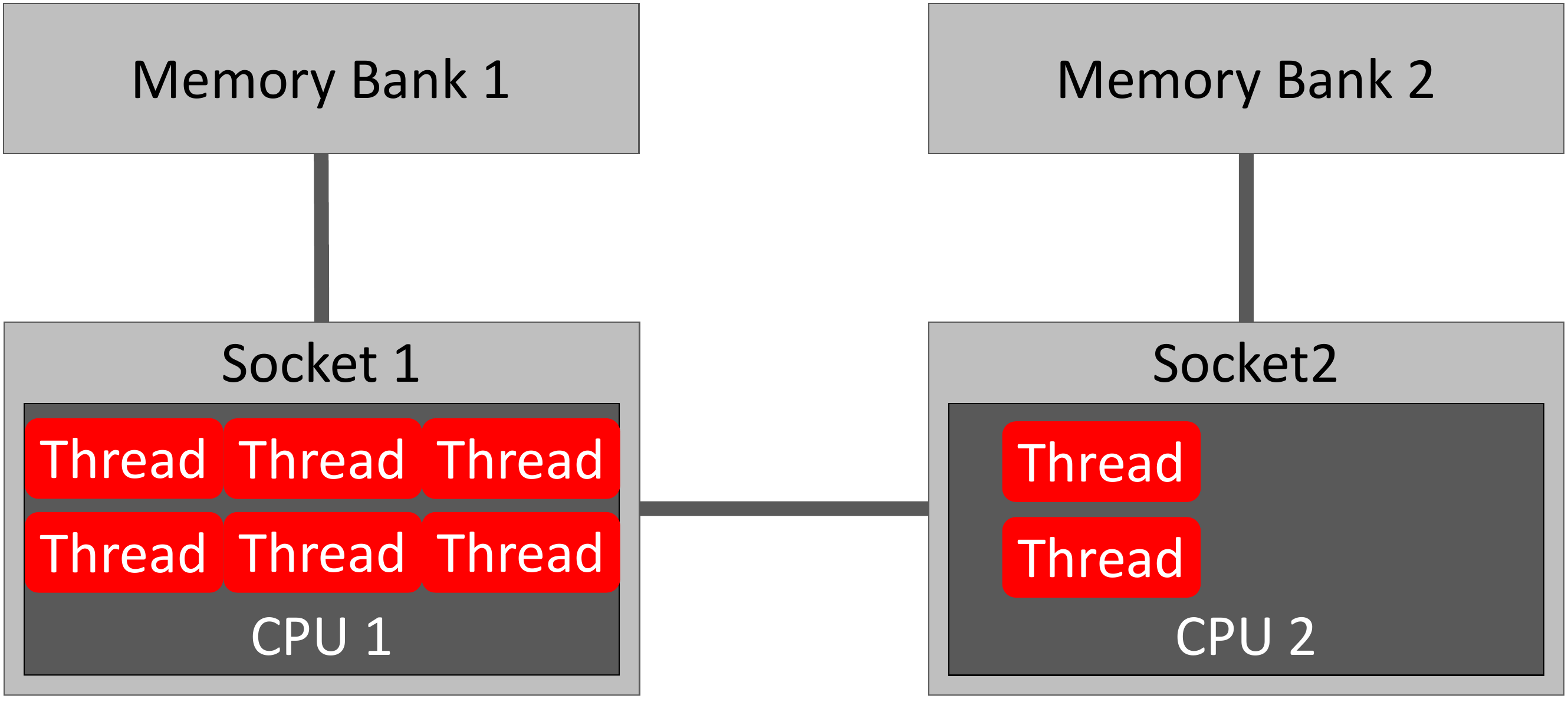}\\
b) Asymmetric\\
\end{center}
\caption{Examples of the two placements we use to determine the types of memory access of an application. These are on a 2 socket machine with 6 cores per socket.}
\label{fig:placements}
\end{figure}

The first of benchmarking runs, is a job with an even number of threads where every thread has its own core, and both sockets have the same thread count. In this placement some cores are left unused to leave space to allow the asymmetric placement to use the same number of threads while maintaining the 1 thread per core policy. The second run uses the same thread count, but has a different number of threads on each socket. An example of these placements can be seen in Figure~\ref{fig:placements}. With our current choice of performance counters, 2 runs is the minimum number from which sufficient data points can be measured to calculate all 8 properties of the signature. While the values for the profiling could be calculated with many different placements the choice to use a symmetric placement for the first run greatly simplifies the process. For example the interleaved and per thread access patterns become identical when the number of threads on each socket is equal, and the thread local accesses place an equal load on each memory bank. Symmetric runs where the symmetry in the loads is not present can be used to detect applications that don't fit the model well. This is discussed in more depth in the evaluation.

To keep profiling times to a minimum it is interesting to note that it is only necessary to execute the workload until a stable state is achieved and the program counters can be read, not until completion. In addition if adding this work into existing performance prediction tools such as Pandia the symmetric run already appears in the existing runs used by the tool, so the asymmetric run is the only additional run.  

\subsection{Data normalization}
The first step is to normalize the data rate per socket relative to the rate of instructions being executed on the socket.  This is required because even on relatively simple workloads there can be a significant variation in execution rate of threads on different sockets. This can be caused by a number of issues including, different latency times on memory accesses, and different available bandwidths to memory banks. By way of example on some lower spec processors the QPI interlink between sockets can be saturated by a single thread, yet to run both the symmetric and asymmetric placements using the same thread count we need at least 2 threads per socket for the symmetric placement. 

\begin{figure}
\begin{center}
\includegraphics[trim=0 152 0 0,clip,width=\linewidth,page=9]{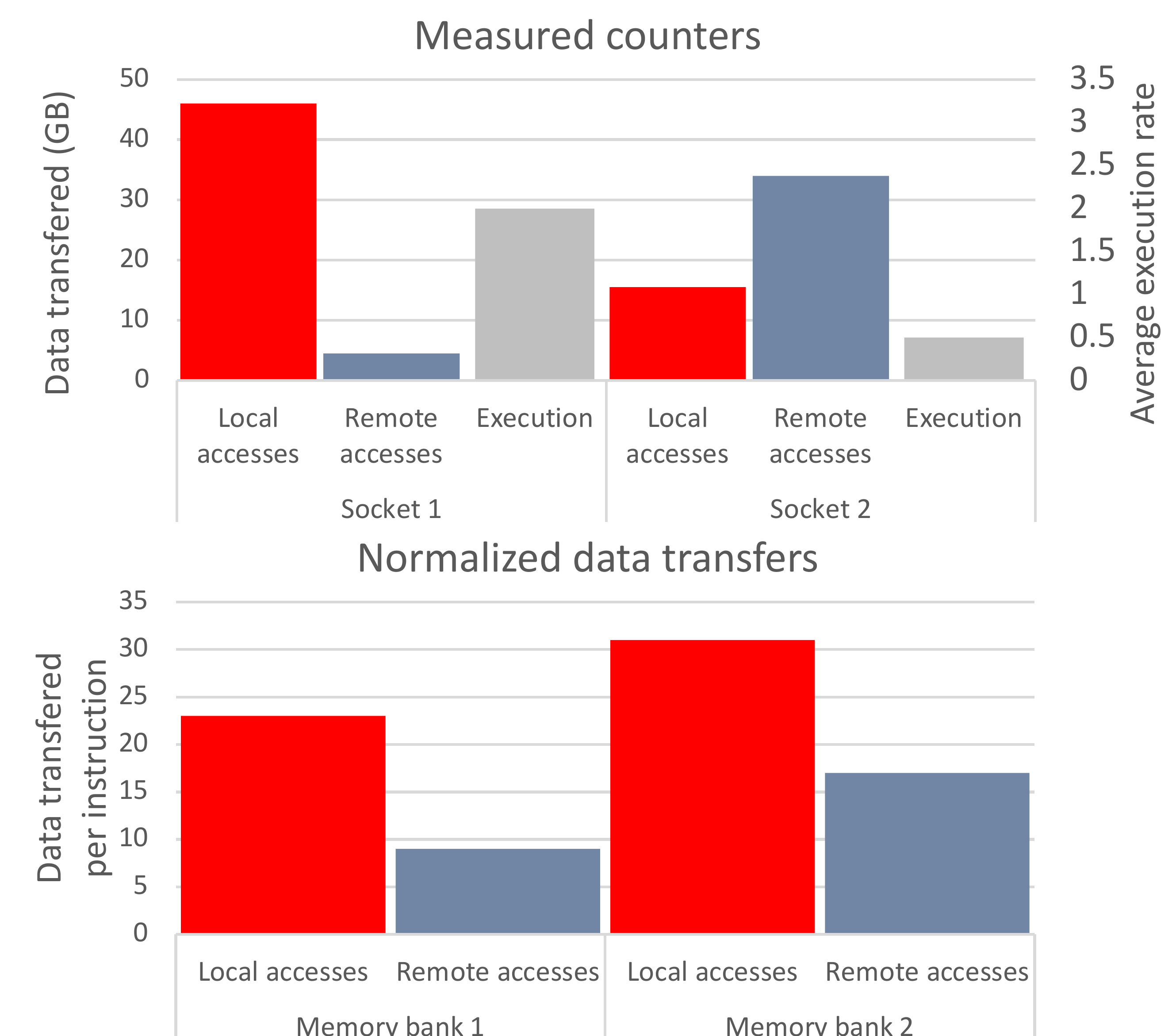}
\end{center}
\caption{Graphs showing an example of the data recorded by the performance counters and the change to this data after normalization. Execution is measured from the perspective of the CPU while memory accesses are from the perspective of the memory bank.}
\label{fig:normalize}
\end{figure}

These differing rates of execution can make the raw counter output unrepresentative of the per thread memory access patterns. For example if we are running the symmetric placement and the threads are performing $\frac{3}{4}$ their accesses locally and $\frac{1}{4}$ their accesses remotely. If all the threads are running at the same speed then both memory banks will report that $\frac{3}{4}$ their accesses are local and $\frac{1}{4}$ are remote, and they each have returned the same amount of data.  However, if the threads on the second socket are running at half the speed of the threads on the first socket then the ratios change so that $\frac{6}{7}$ of the accesses to bank 1 and $\frac{6}{10}$ of the accesses to bank 2 are local. Whats more bank 1 only gets $\frac{7}{8}$ and bank 2 gets $\frac{5}{8}$ of the traffic they would receive if the threads ran at full speed.

To overcome this we first normalize the data transfer counters to the rate that the threads are executing at a per socket granularity. The resulting output is the data sent or received per average instruction execution rate per memory bank from each socket . An example can be seen in Figure~\ref{fig:normalize}. To do this for each memory bank we record the remote reads, the remote writes, the local reads, and the local writes. Each of these is then divided by the average rate instructions were executed by the threads on the socket that that traffic was to or from. 

\subsection{Static fraction}
\begin{figure}
\begin{center}
\includegraphics[trim=0 717 0 0,clip,width=\linewidth,page=11]{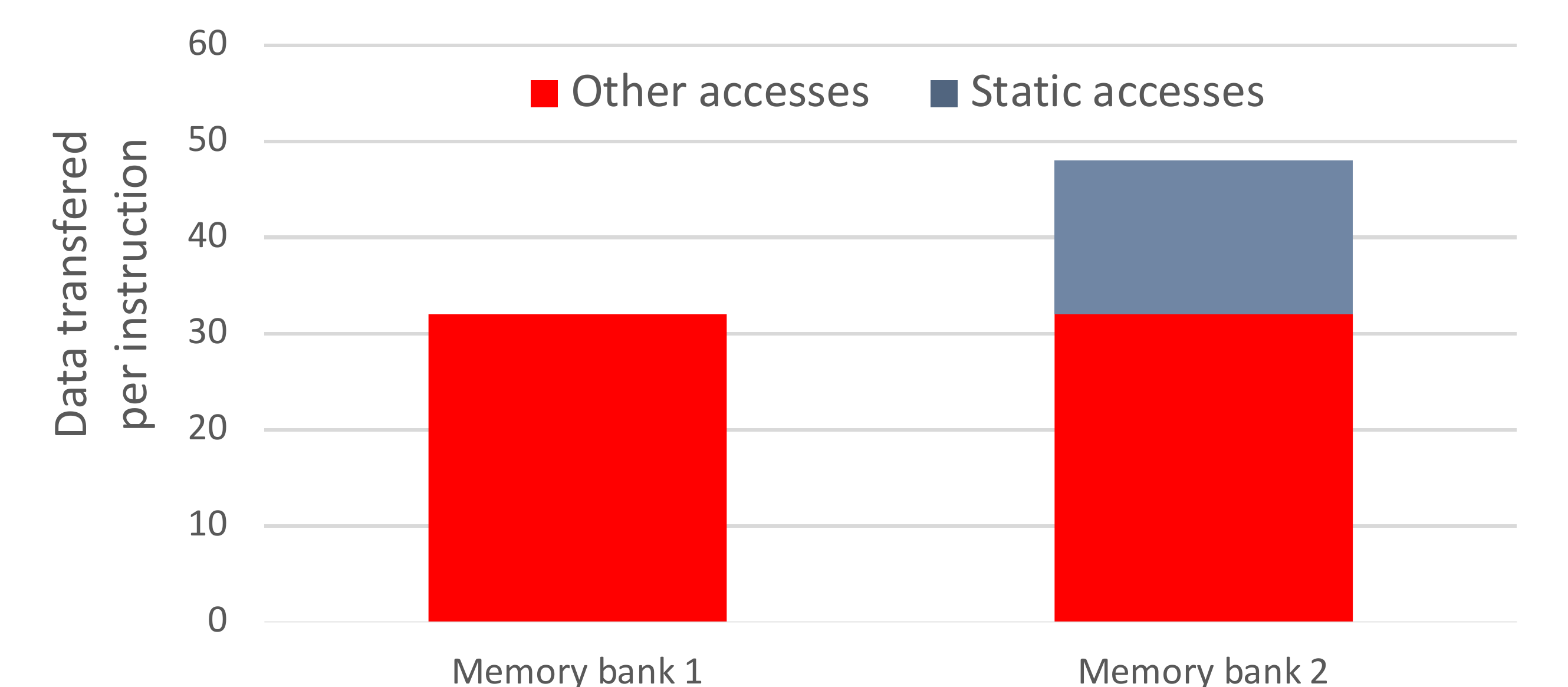}
\end{center}
\caption{Total bandwidth per socket, with the data associated with accesses to static allocations highlighted.}
\label{fig:static}
\end{figure}

After normalizing the data transfers the first properties calculated for the job description are the static socket, and the static fraction. To calculate these we use the normalized results of the symmetric profiling run. For each memory bank we sum the local and remote reads~($\text{l\_reads}$ and $\text{r\_reads}$) or writes to generate the total normalized reads or writes to the memory bank. For example:
$$\text{reads}_\text{bank 1} = \text{l\_reads}_\text{bank 1} + \text{r\_reads}_\text{bank 1}$$
From these we first calculate the socket that the static fraction is associated with by observing which memory bank transfered the largest volume of data, in the case of the example in Figure~\ref{fig:normalize} this is socket 2 as shown in Figure~\ref{fig:static}.

Having determined the static socket we next calculate the static fraction. To do this we calculate the additional data transfer on the static socket relative to the other sockets and divide this additional transfer by the total transfer used by the workload. This gives us the fraction of the transfer that was for data statically allocated to a single socket, and there in the fraction of the bandwidth that will be to the static socket for each thread.

$$
\text{static fraction} = \frac{\text{reads}_\text{bank 2} - \text{reads}_\text{bank 1}}{\text{reads}_\text{bank 1} + \text{reads}_\text{bank 2}}
$$

\vspace{.4cm}

\noindent In this example this works out as 0.2 or 20\%.

\subsection{Local fraction}

To calculate the fraction of data transfers due to accesses to thread local data we first remove the data transfer due to the static fraction. In the working example this is just a case of deducting half the bandwidth associated with the static fraction from bank 2's remote accesses and half from its local accesses. 

\begin{figure}
\begin{center}
\includegraphics[trim=0 722 0 0,clip,width=\linewidth,page=12]{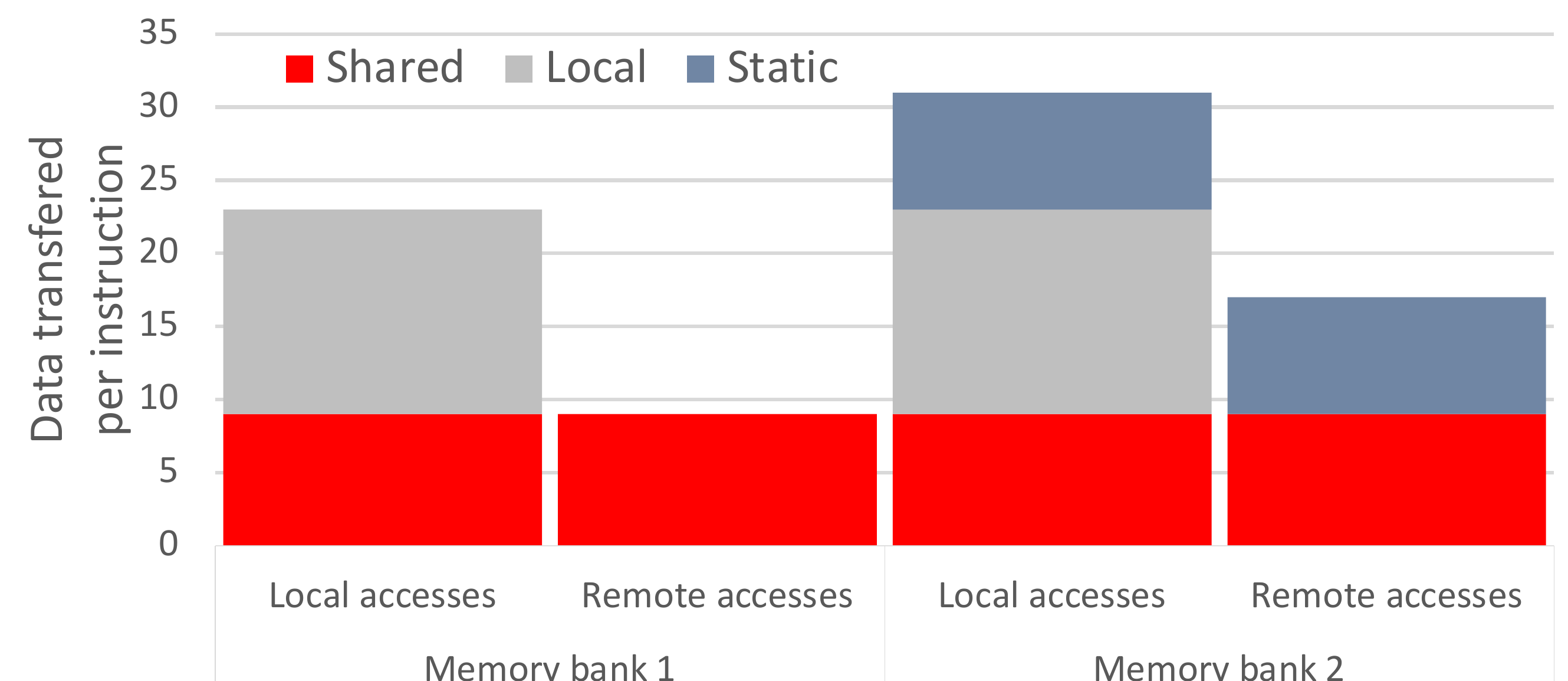}
\end{center}
\caption{Remote and local accesses to memory banks on each socket. The accesses already classed as static and removed from our calculations are highlighted along with the accesses that will be classes as thread local.}
\label{fig:local}
\end{figure}

This leaves us with the accesses to data shared between sockets which is made up of per thread data and interleaved data, and the accesses to data that is only used by threads on a given socket, local data. For each socket we then calculate the fraction of accesses that are remote. 
$$r_i=\frac{\text{remote accesses}_{\text{bank}_i}}{\text{remote accesses}_{\text{bank}_i} + \text{local accesses}_{\text{bank}_i}}$$
 For the symmetric pattern Interleaved and Per thread accesses are indistinguishable meaning we can deduce that if there were no accesses to local memory allocations, for $s$ sockets we would expect the fraction of remote accesses to be $r = \frac{s-1}{s}$. Adding back in the possibility of a non zero local fraction including scaling to allow for the static fraction that we have already removed from the bandwidths gives: 
$$r = \frac{s-1}{s}\left(1-\frac{\text{local fraction}}{(1-\text{static fraction})}\right)$$ 
This can be rearranged to get the local fraction. In our example the measured value of $r$ is $0.28125$, with no bandwidth categorized as local memory accesses we would expect $r=0.5$ in this example as $s=2$. From this we calculate that the local fraction is $0.35$. This split along with the static accesses can be seen in Figure~\ref{fig:local}.

\subsection{Per thread fraction}
\begin{figure}
\begin{center}
\includegraphics[trim=0 722 0 0,clip,width=\linewidth,page=13]{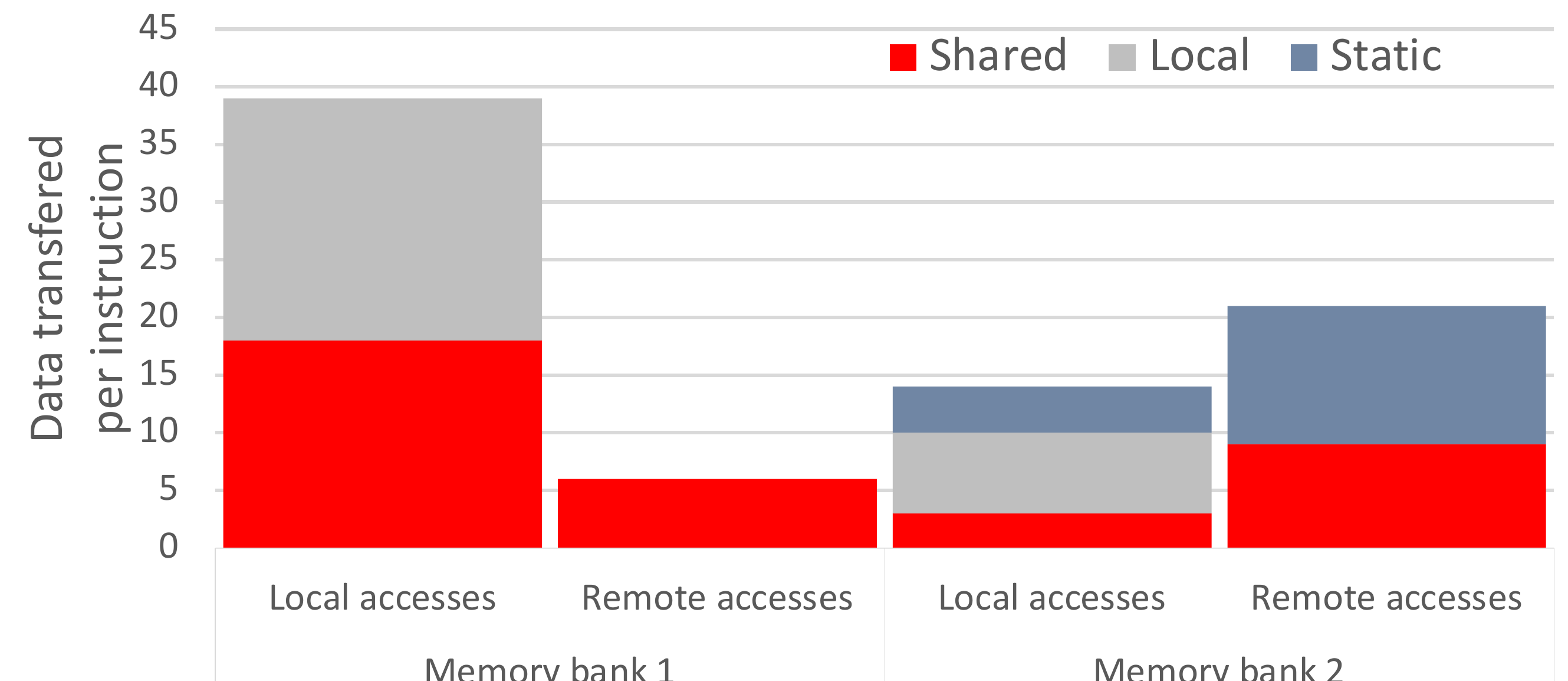}
\end{center}
\caption{The normalized results for the asymmetric case with the static and local components calculated and removed from the computation of the per thread fraction.}
\label{fig:removingKnowParts}
\end{figure}

As discussed on a symmetric placement per thread and interleaved accesses are indistinguishable. To overcome this we use a run with an asymmetric placement to calculate this value. For our worked example we will use the placement in Figure~\ref{fig:placements}. Taking the results for this new placement we sum the bandwidth for each CPU, for example for reads: 
$$\text{reads}_\text{CPU 1} = \text{l\_reads}_\text{bank 1} + \text{r\_reads}_\text{bank 2}$$
Next we remove from the static socket the static fraction of the bandwidth. In our example this is done for reads as follows:
$$ \text{r\_reads'}_\text{bank 2} = \text{r\_reads}_\text{bank 2} - \text{static fraction} \times \text{reads}_\text{CPU 1} $$
$$ \text{l\_reads'}_\text{bank 2} = \text{l\_reads}_\text{bank 2} - \text{static fraction} \times \text{reads}_\text{CPU 2} $$
We then remove the fraction of the local bandwidth associated with each memory bank. This is done as follows: 
$$ \text{l\_reads'}_\text{bank 1} =  \text{l\_reads}_\text{bank 1} - \text{local fraction} \times \text{reads}_\text{CPU 1} $$
$$ \text{l\_reads''}_\text{bank 2} = \text{l\_reads'}_\text{bank 2} - \text{local fraction} \times \text{reads}_\text{CPU 2} $$
 The result of this for our worked example can be seen in Figure~\ref{fig:removingKnowParts}.

Having removed the already accounted for elements of the bandwidth we calculate for each CPU the fraction of its bandwidth that is used for transfers to its local memory bank~($l$). 

$$l=\frac{\text{local accesses}_{\text{bank}_i}}{\text{local accesses}_{\text{bank}_i} + \sum_{j=1}^{s j \neq i}\text{remote accesses}_{\text{bank}_j}}$$  
 In our worked example these are $\frac{2}{3}$ and $\frac{1}{3}$ for sockets 1 and 2 respectively.

We also calculated the expected value of~$l$ if all the data is accessed on a per thread basis. This value is given by:
$$
\text{Per thread data}_i = \frac{n_i}{\sum_{j=1}^{s}n_j}
$$

\noindent where $n_i$ is the number of threads on socket $i$ and $s$ is the number of sockets. Next we calculate the expected local fraction if all the data is interleaved:

$$
\text{Interleaved}_i = \frac{1}{s}
$$

In our example this provides fractions of $\frac{3}{4}$ and $\frac{1}{4}$ if all bandwidth is to Per thread data and $\frac{1}{2}$ and $\frac{1}{2}$ if all bandwidth is to Interleaved data. As the combination for a given application will be somewhere between these two points we can interpolate between them:

$$
\parbox{.7cm}{\begin{center}$l_i$\end{center}} = \parbox{1.7cm}{\begin{center}Per thread data$_i$\end{center}} \times p + \parbox{1.7cm}{\begin{center}Interleaved data$_i$\end{center}} \times (1-p)
$$  

This equation can then be rearranged to get the value of $p$, in out example $\frac{2}{3}$. $p$ can then be scaled to get the Per thread fraction as follows:
$$
\text{per thread fraction} = p \times (1 - \text{local fraction} - \text{static fraction})
$$
 In the worked example this is 0.3. The Per thread fraction is bounded between $[0\ldots1]$ to ensure that unusual data patterns cannot cause unexpected effects. We will discuss this further in Sections~\ref{sec:evaluation} and~\ref{sec:futureWork}.

\section{Evaluation}
\label{sec:evaluation}
We describe the evaluation in 3 sections. First we examine synthetic applications where we have control over the memory placement and compare the measured placement with the known values to confirm the model is successfully identifying the access patterns. Second we look at the signatures generated for a set of benchmarks designed to mimic real world applications. We compare these on the different hardware to examine the stability of the model with more complex workloads. Finally we compare the prediction of the model using these signatures with measured results to gauge the accuracy of the model.

Two experimental systems were used to evaluate this work. These systems are dual socket machines populated with Xeon E5-2630 v3, and Xeon E5-2699 v3 processors which are 8 core, and 18 core Haswell processors respectively. While the processor architectures are similar, the communications profile of the system as a whole is very different as shown in the Figure~\ref{fig:commuunicationSpeed}. From this we can see that both systems have similar read and write bandwidths to local memory, but drastically different performance when accessing remote memory where the 8 core processors only have 0.16 of the bandwidth for remote reads and 0.23 of the bandwidth for remoter writes  relative to local reads and writes. On the 18 core processors are the bandwidths are much more comparable to local bandwidths with 0.59 of the bandwidth for remote reads and 0.83 of the bandwidth for remote writes.

As Linux may try to adjust the placement of memory during the execution of the benchmark we disable autonuma in all our tests. This ensures we measure the benchmarks in a stable state.

\subsection{Synthetic benchmarks}
The synthetic benchmarks perform index chasing through an array. Arrays of integers are constructed such that each element in the array is an index to the next element that should be read i.e. at each step \texttt{i = A[i]}. For these tests the indexes access elements sequentially with a stride size of a cache line and the final element indexing the first. The arrays are typically in the order of gigabytes in size. This means they fit in memory, but not in the cache. This results in a regular access pattern that hardware prefetchers can use to improve performance, minimal use of the cache so we have the strongest signal to noise ratio possible, and code that cannot be analyzed and optimized away at compile time as the content of the array is unknown to the compiler.  For Static,  Local, and Interleaved each thread constructs its own loop in an array and iterates round it. In the cases of Static and Interleaved we used \texttt{numactl} to enforce the placement of the memory, while local relies on a first touch memory policy to ensure the required pages are placed local to the requesting thread. For Per thread each thread constructs an array, and each thread will then iterate through each array in turn. 

\begin{figure}
\begin{center}
\includegraphics[trim=0 712 0 0,clip,width=\linewidth,page=10]{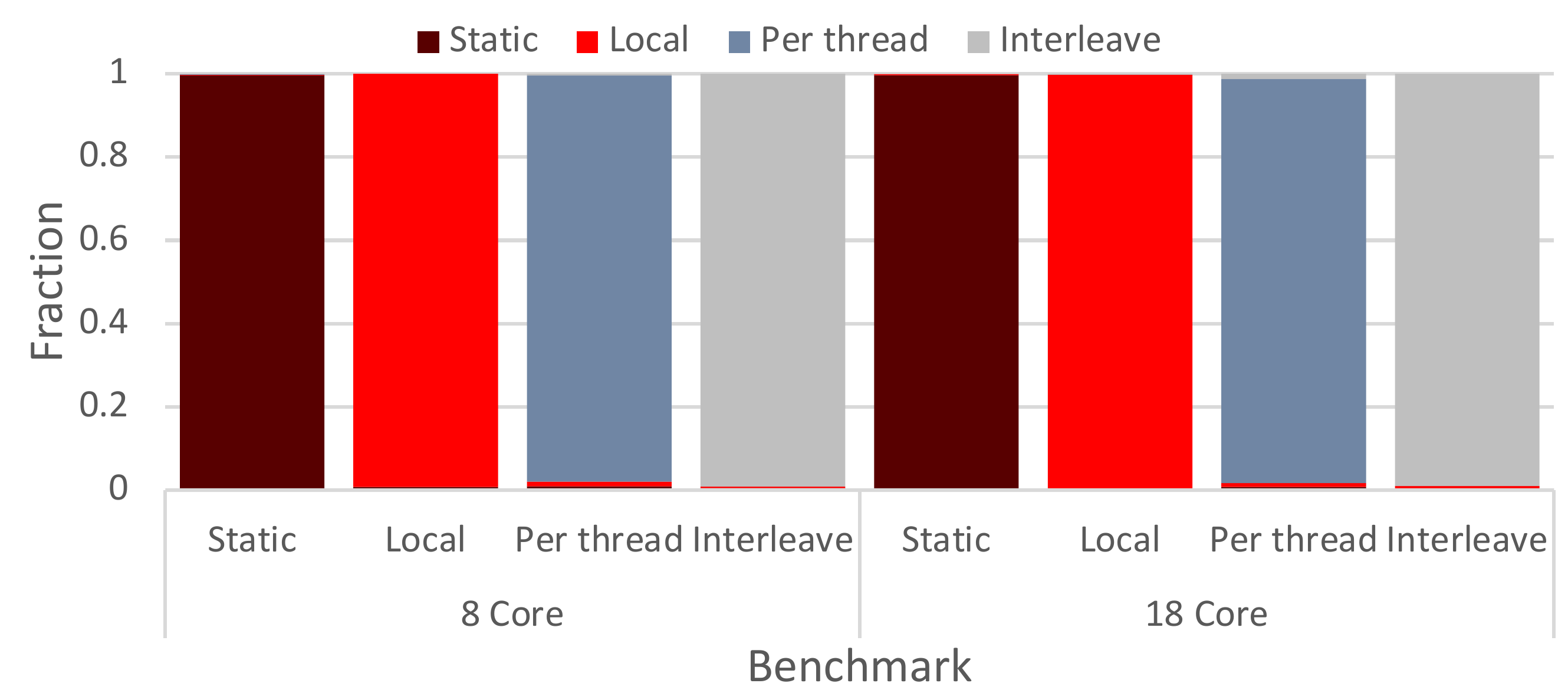}
\end{center}
\caption{Graphs of the memory signature measured for each of the synthetic benchmarks on 8 and 18 core Haswell processors.}
\label{fig:synthetic}
\end{figure}

The results from profiling these benchmarks can be seen in Figure~\ref{fig:synthetic}. For all the benchmarks the results are as expected with the largest volume of miscategorized bandwidth measuring less than 0.9\% which is spread evenly across the remaining 3 categories and can easily be accounted for by background noise.

\subsection{Real benchmarks}
The workloads for these tests are drawn from a range of sources: The NAS parallel benchmark suite (NPB)~\cite{NPB}; The SPEC OpenMP workloads (OMP)~\cite{SpecOpenMP}; In-memory graph analytics workloads (GA) from Harris \emph{et al}~\cite{harris:atc}; And database join operators (DBJ) from Balkesen \emph{et al}~\cite{balkesen:main}. They are designed to mimic real world applications and we describe them in Table ~\ref{tab:benchmarks}.
\vspace{.2cm}

\begin{table}
\begin{tabular}{@{}|l|p{6.1cm}|}
\hline
\textbf{Benchmark} & \textbf{Description} \\
\hline
Applu & Parabolic / Elliptic PDE solver (OMP)\\
Apsi& Meteorology pollutant distribution (OMP)\\
Art & Neural network simulation (OMP)\\
BT & Block tri-diagonal solver (NPB)\\
Bwaves & Blast wave simulation (OMP)\\
CG & Conjugate gradient (NPB)\\
EP & Embarrassingly parallel (NPB)\\
Equake & Earthquake simulation (OMP)\\
FMA-3D & Finite-element crash simulation (OMP)\\
FT & Discrete 3D fast Fourier transform (NPB)\\
IS & Integer sort (NPB)\\
LU & Lower-upper Gauss-Seidel solver (NPB)\\
MD & Molecular dynamics simulation (NPB)\\
MG & Multi-grid on a sequence of meshes (NPB)\\
NPO & No partitioning, optimized hash join (DBJ)\\
PRHO & Parallel radix histogram optimized hash join (DBJ)\\
PRH & Parallel radix histogram hash join (DBJ)\\
PRO & Parallel radix optimized hash join (DBJ)\\
Page rank & In-memory parallel Page rank (GA)\\
Sort join & In-memory sort-join (DBJ)\\
SP & Scalar Penta-diagonal solver (NPB)\\
Swim & Shallow water modeling (OMP)\\
Wupwise & Wuppertal Wilson fermion solver (OMP)\\
\hline
\end{tabular}
\caption{Description of benchmarks.}
\label{tab:benchmarks}
\end{table}

\begin{figure}
\begin{center}
\includegraphics[trim=0 152 0 0,clip,width=\linewidth,page=18]{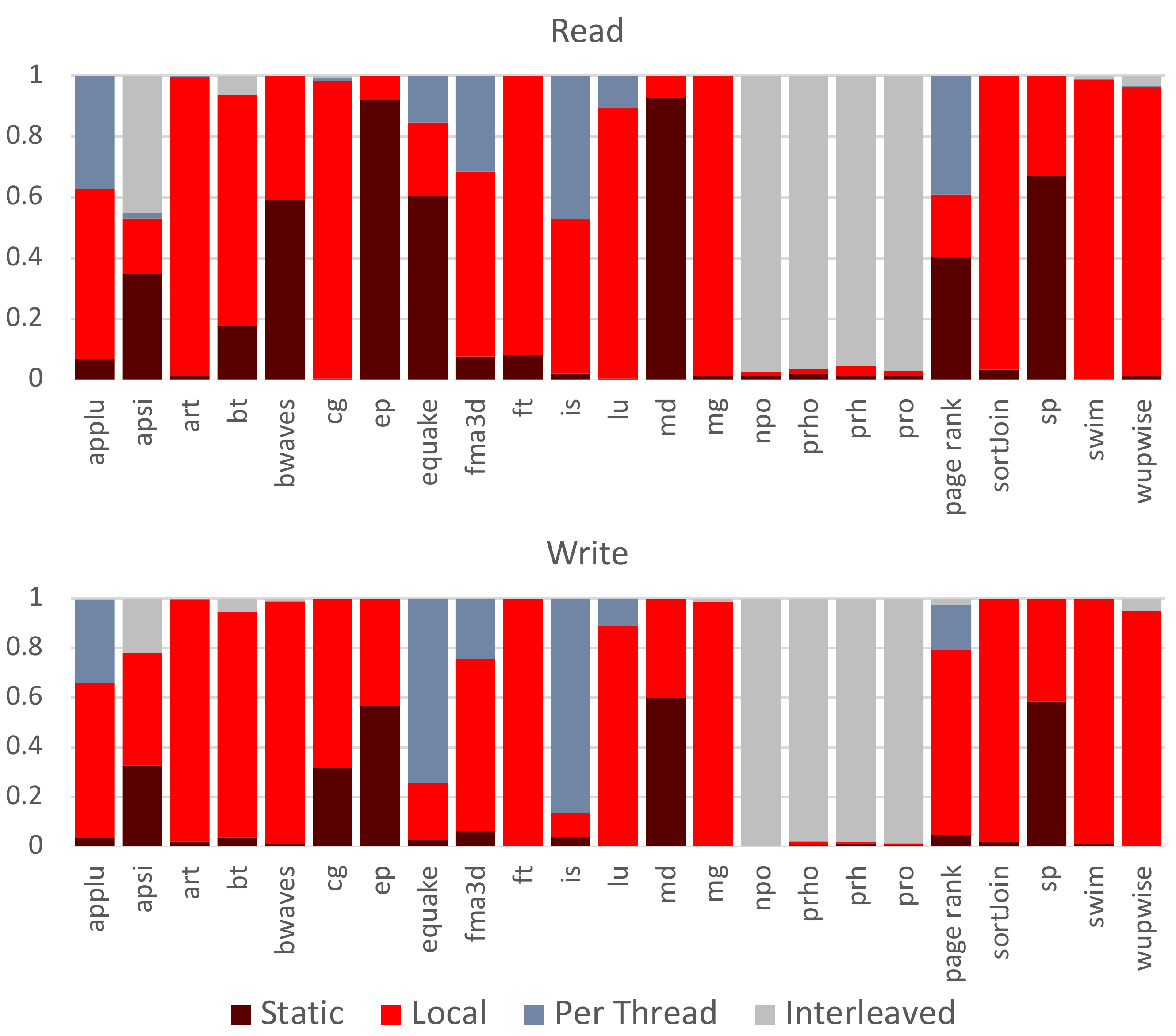}
a) 8 core machine
\includegraphics[trim=0 152 0 0,clip,width=\linewidth,page=17]{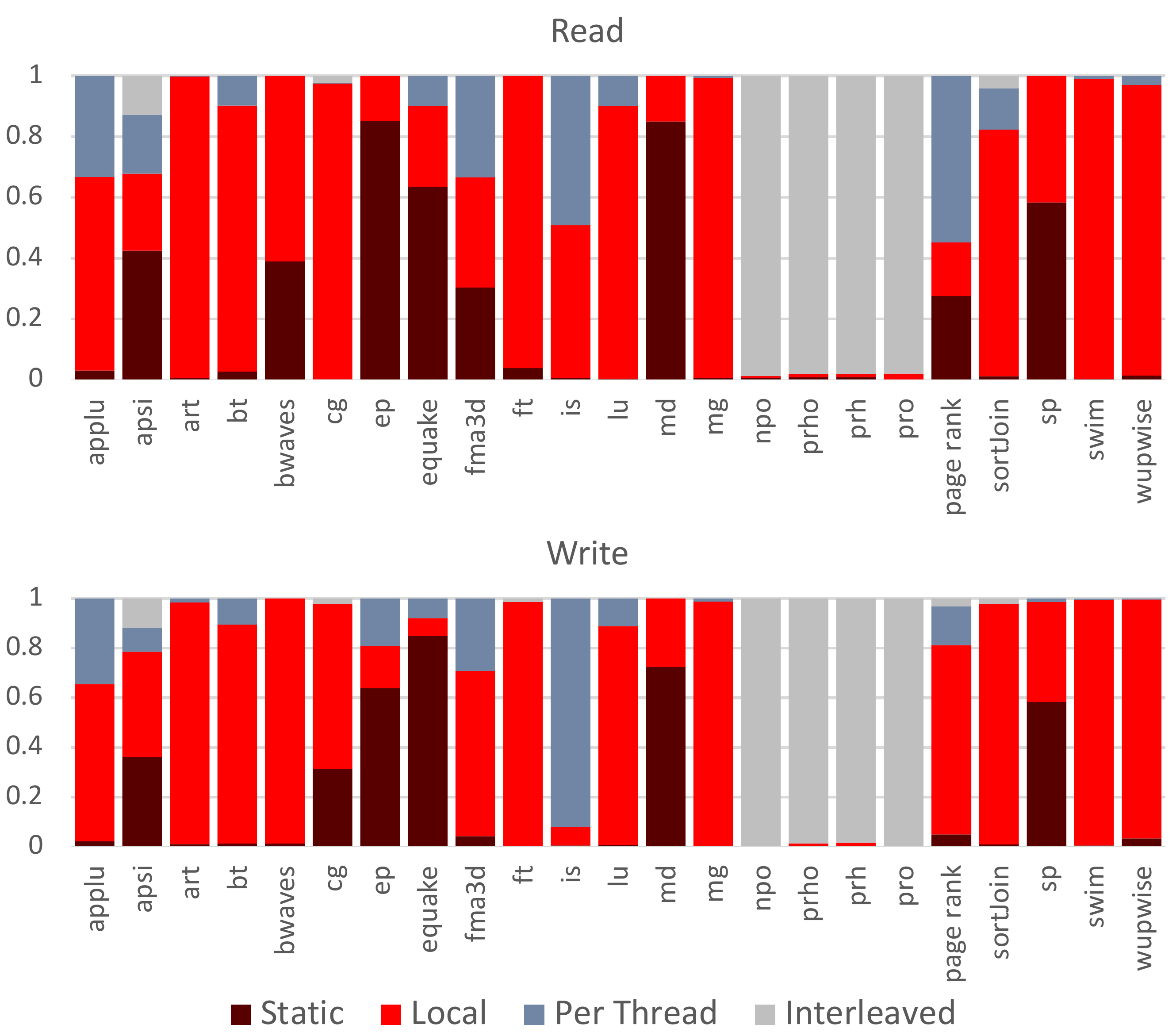}
\end{center}
b) 18 core machine
\caption{Graph of the bandwidth signatures generated by the benchmarks for reading and writing on the different systems}
\label{fig:benchmarkSignatures}
\end{figure}

\subsubsection{Model stability}
\begin{figure}
\begin{center}
\includegraphics[trim=0 712 0 0,clip,width=\linewidth,page=19]{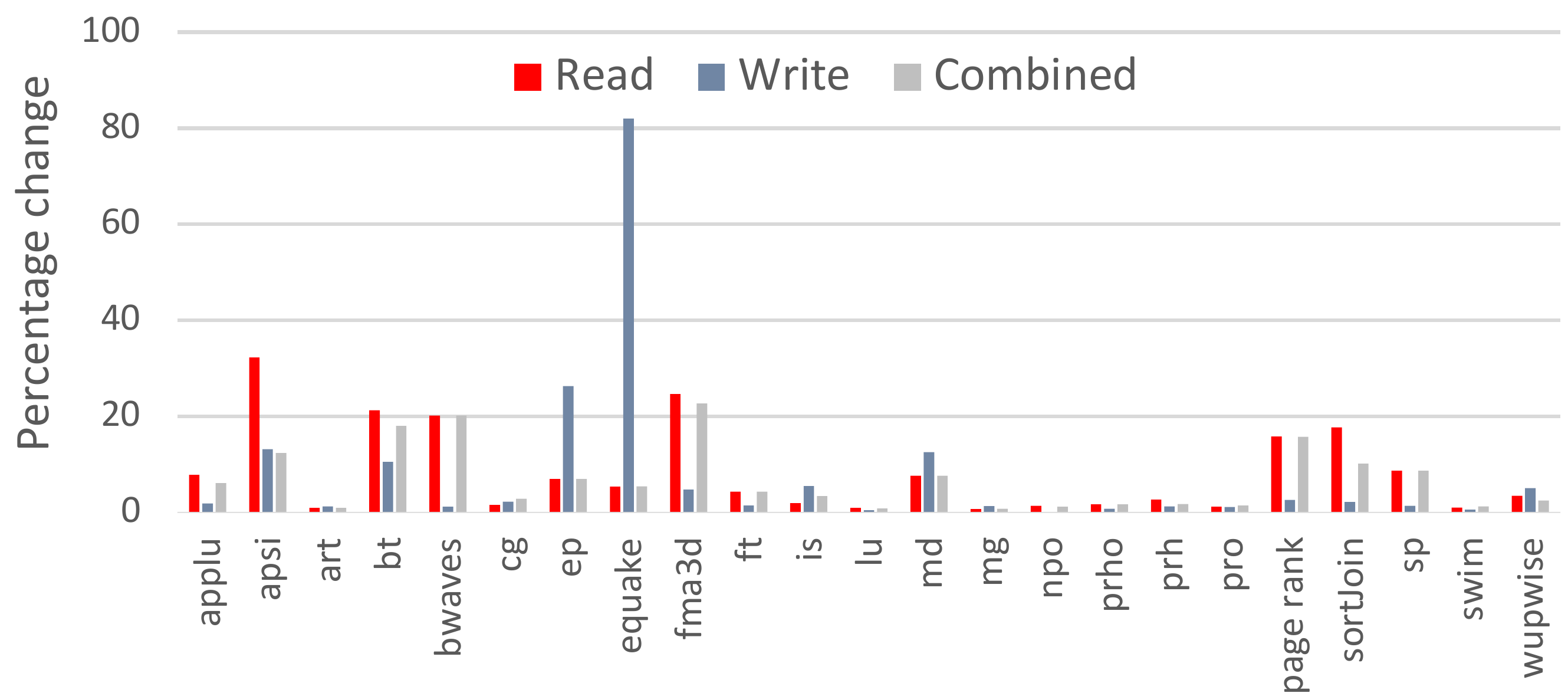}
\end{center}
\caption{Graph of the percentage change in the bandwidth placement between systems.}
\label{fig:signatureChanges}
\end{figure}

\begin{figure}
\begin{center}
\includegraphics[trim=0 712 0 0,clip,width=\linewidth,page=21]{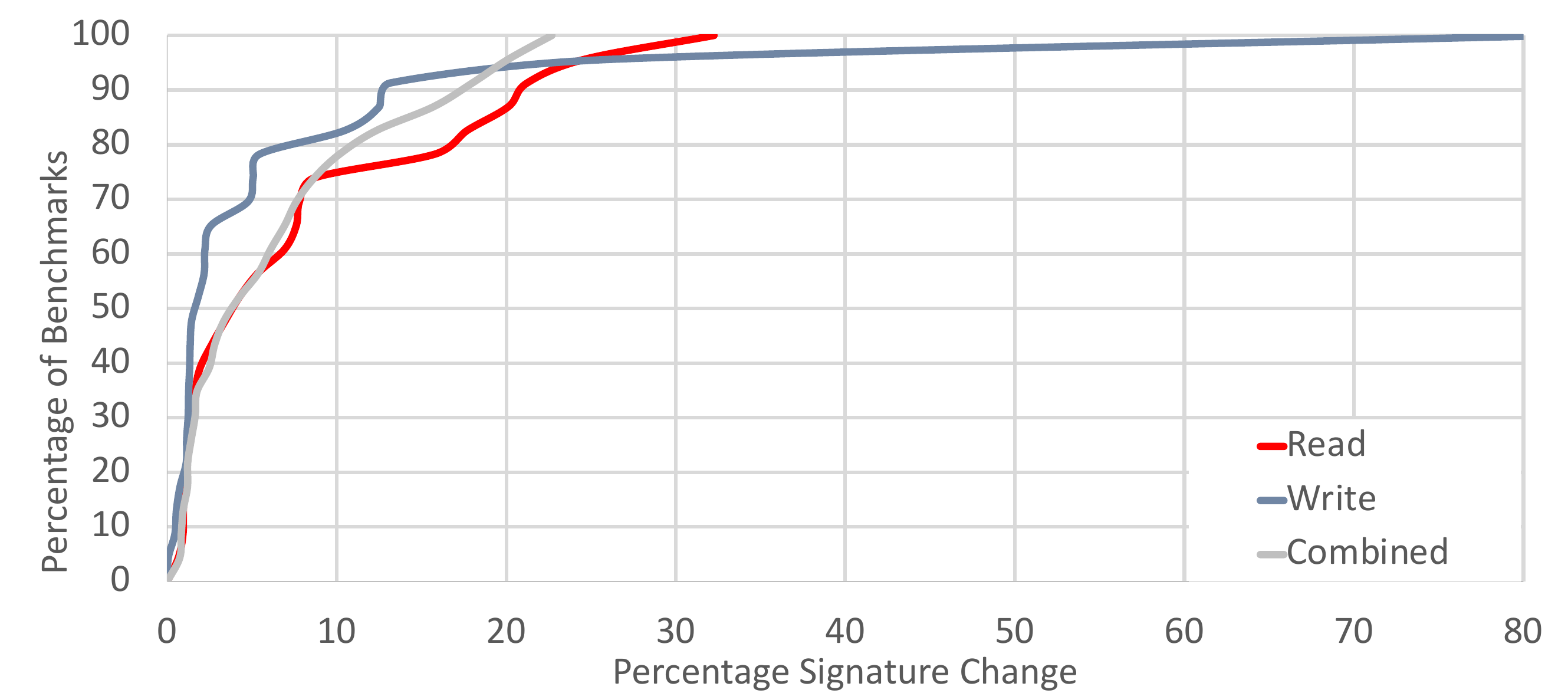}
\end{center}
\caption{Cumulative frequency graph of the change in the benchmarks signatures relative to the percentage of the benchmarks.}
\label{fig:cumulative}
\end{figure}

Calculating the bandwidth description on both machines for each benchmark for both reads and writes we get the signatures in Figure~\ref{fig:benchmarkSignatures}. Figure~\ref{fig:signatureChanges} compares the percentage of the bandwidth that is reallocated between the two different signatures. At first glance some of these results appear to be very bad with a change in excess of 80\% for equake writes. This is due to this benchmark performing almost exclusively reads with the very small number of writes resulting in a very low signal to noise ratio. So while the signature does change for these examples the bandwidth associated with this is negligible. To illustrate this we also plot the difference in signature value if instead of constructing a separate signature for reads and writes, we combine the bandwidths and construct the signatures using these combined bandwidth figures. The combined figures for equake change by 5.4\%. The mean change for all the benchmarks 6.8\% and the median is 4.2\%. 

\begin{figure}
\begin{center}
\includegraphics[trim=0 712 0 0,clip,width=\linewidth,page=23]{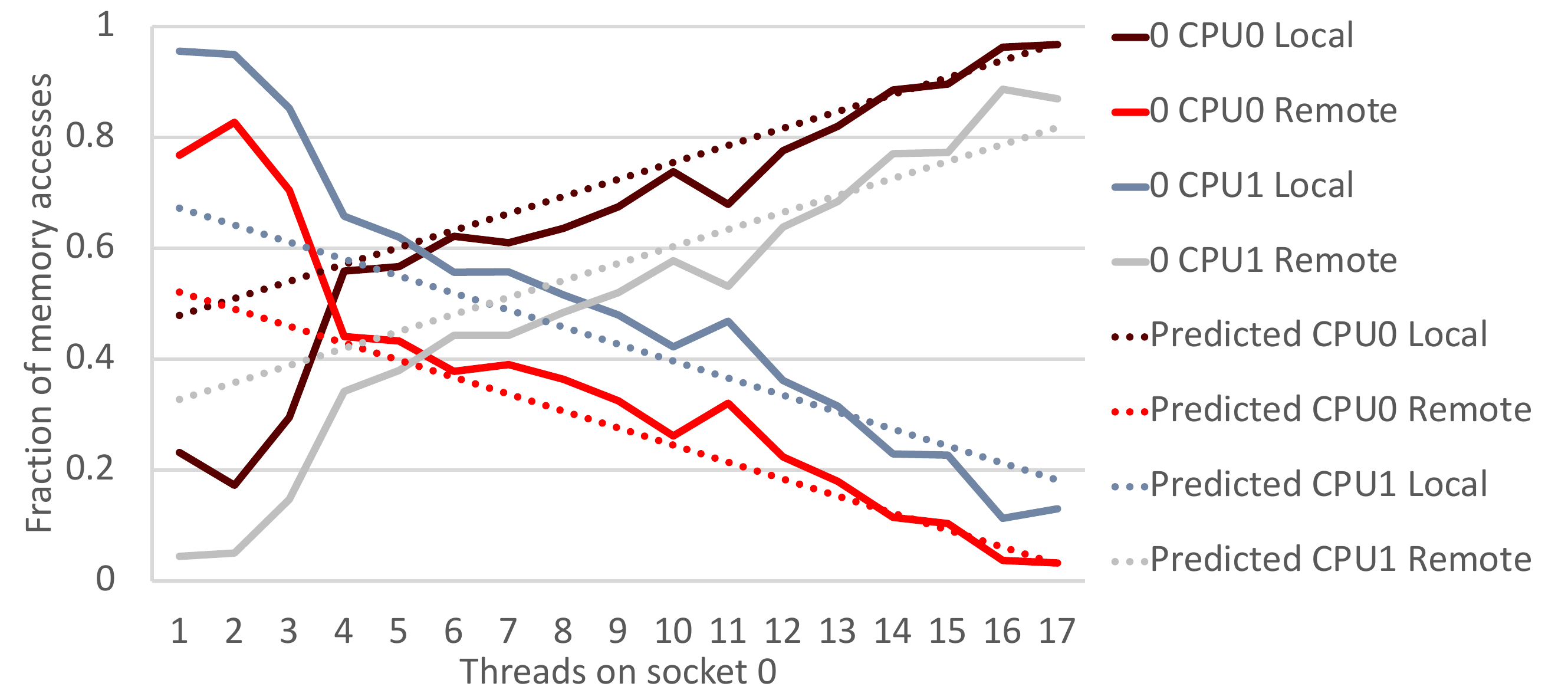}
\end{center}
\caption{Graph of the measured and predicted results for the combination of reads and writes with Page rank.}
\label{fig:pagerankResults}
\end{figure}

The difference between the mean and median highlights that while for most of the benchmarks the signature is stable and usable for a small number there are significant errors. This can be seen inf Figure~\ref{fig:cumulative} where we show a cumulative frequency graph of the benchmarks and the change between signatures. From this we can see that over 50\% of applications have a change of less than 5\% and over 75\% of applications have a change of less than 10\%. However, there is a set of benchmarks for which the model does not provide a very good fit to the access patterns. These occur when the bandwidth requirements varies between threads and is especially problematic when it changes with the number and position of the threads. For example in the Page rank benchmark the nodes in the graphs are listed in the order they were visited when the dataset was collected. As a result of being collect through walking the graph well connected nodes are more likely to be met first and so appearing earlier in the dataset. This happens because after a short period of exploring the graph the walker may reach a well connected segment. From here it will probably mostly reach other well connected nodes. The later nodes to be reached will more likely be weakly connected hence taking longer to be discovered by the walker. As a result the part of the graph that appears earlier in the dataset is better connected on average than the rest of the graph. This results in higher local bandwidth requirements on the first socket which will erroneously be marked as static bandwidth. This then confuses the calculation of both local and per thread fractions. When the threads are moved around, the bandwidth requirements fail to change in the way described. An example of this can be seen in Figure~\ref{fig:pagerankResults}.

Fortunately it is possible to detect when situations like this occur as there is redundant information in the program counters that highlights the inconsistency.  For example once we remove the static fraction with the symmetric placement we expect the placement to be symmetric. If when we examine the local remote ration for each socket we find that it is not symmetric this is a sign that the application does not fit the model. The bigger the difference the worse the fit. 

\subsubsection{Model accuracy}
To test the accuracy of the model with the different benchmarks we executed each benchmark with the largest thread count that it could support on a single socket with at most one thread per core. For each benchmark configuration we then varied the distribution of the threads between the two sockets maintaining a single thread per core. Measuring the local and remote reads and writes for each socket and comparing against the read, write, and combined model predictions gives a large number of comparison points with the 18 core machine providing 2322 data points to compare the predicted results against. The results of this for the Page rank benchmark when comparing the combined reads and writes bandwidth prediction can be seen in Figure~\ref{fig:pagerankResults}. This shows the discussed poor fitting of the model to threads working on the first section of the graph, and the more effective modeling of the bandwidth requirements for processing the rest of the graph.

\begin{figure}
\begin{center}
\includegraphics[trim=0 712 0 0,clip,width=\linewidth,page=26]{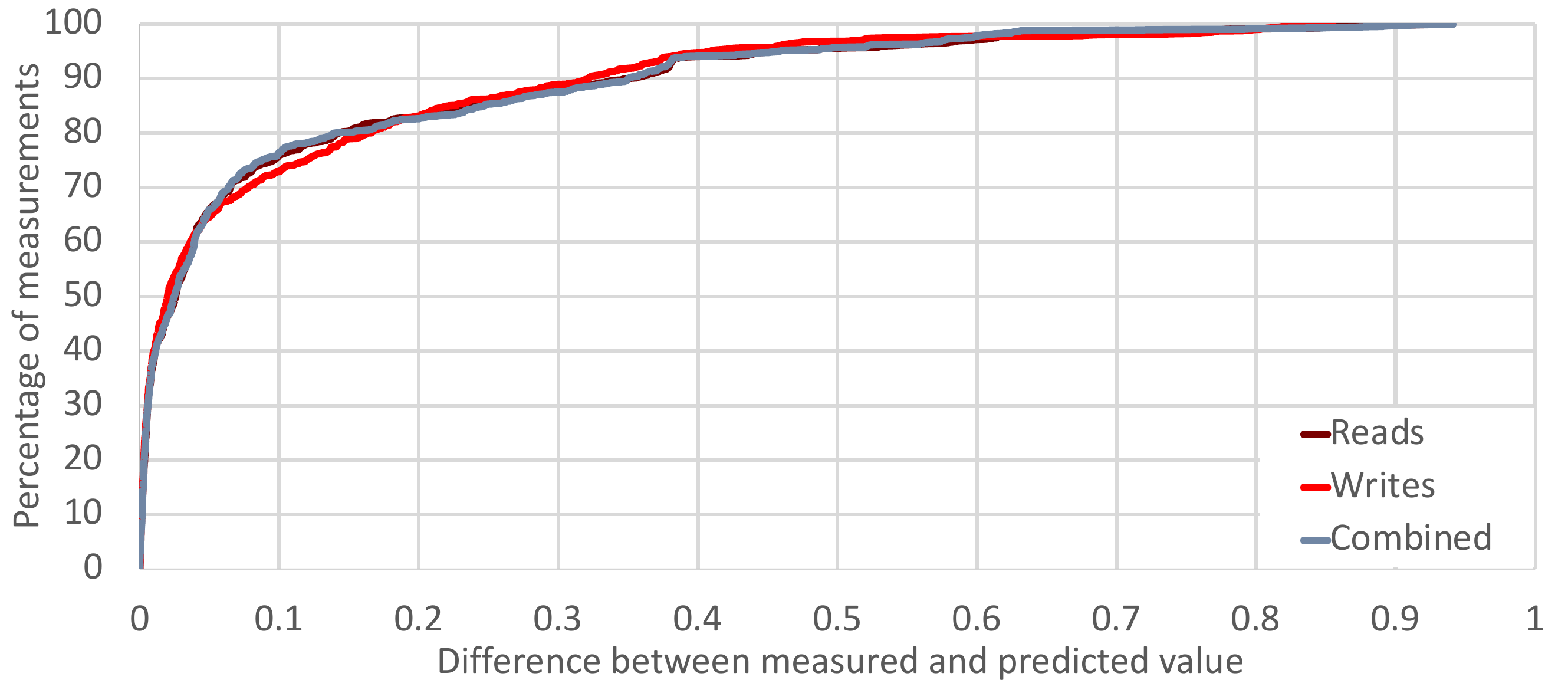}
\end{center}
\caption{Cumulative frequency graph of the percentage of the benchmark measurements vs the size of the error.}
\label{fig:cumulativeError}
\end{figure}

\begin{figure}
\begin{center}
\includegraphics[trim=0 712 0 0,clip,width=\linewidth,page=25]{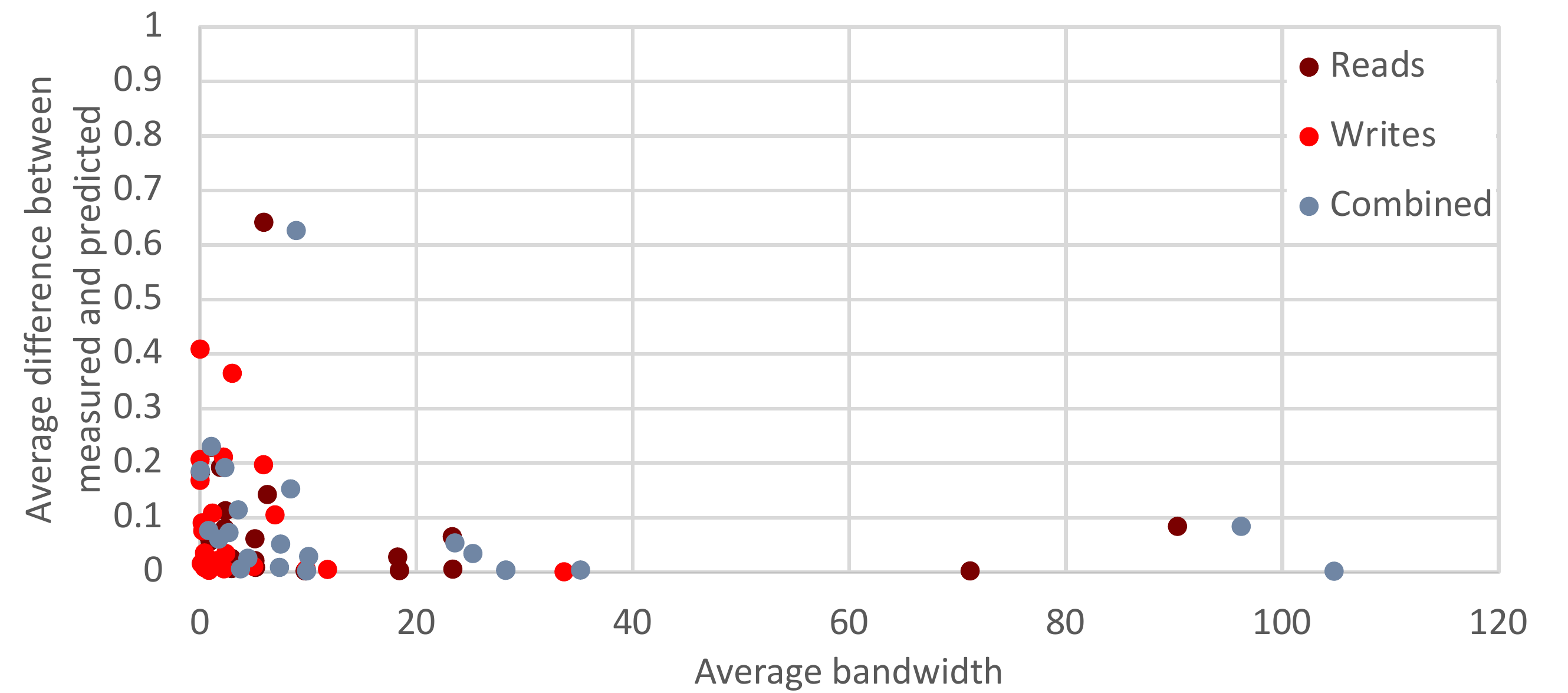}
\end{center}
\caption{Graph of the average size of the error vs the average amount of bandwidth required by the benchmark.}
\label{fig:bandwidthVsError}
\end{figure}

For each data point we record the difference between the measured value and the predicted value. Looking at these across all our experiments we can generate the cumulative frequency graph in Figure~\ref{fig:cumulativeError}. This shows that for over 50\% of the measurements the difference is less than 2.5\% of the total bandwidth, and for 75\% of measurements the difference is less than 10\% of the total bandwidth. Looking at where these differences occur in Figure~\ref{fig:bandwidthVsError} we plot the average difference for each benchmark relative to the average bandwidth used by the benchmark. This shows that the substantial errors only occur in the benchmarks with low bandwidth requirements. Such benchmarks both have a low signal to noise ratio, but also do not require high accuracy as by definition they are not moving large quantities of data so will be less affected, and less likely to interfere with the data transfers of other workloads.

\section{Future work}
\label{sec:futureWork}
As mentioned in the evaluation the current model has one key limitation. It assumes that each thread accesses data with the same frequency relative to its rate of execution. However as described there is information that allows for these situations to be detected. It would be preferable to instead of just detecting these situations to model them as well, extending the applicability of this model. This is a focus for future work.

\section{Related work}
Asymmetry and variance of bandwidth performance on NUMA nodes leading to contention on resources has long been identified as a problem~\cite{Blagodurov2011, Merkel2010, Zhuravlev2010}. This has resulted in the development of many techniques to improve the placement of threads and memory.  These pieces of work measure the performance of the NUMA machine or the loads being executed through a range of techniques including: 
\begin{itemize}
\item Monitoring performance counters while executing either known benchmarks, or arbitrary workloads~\cite{Blagodurov2011,Merkel2010,Majo2011}. The performance counters monitored are typically LLC misses and instructions executed, but can be far more extensive~\cite{Goodman:2017:pandia, smartArrays}, and extend to applying ML techniques to every counter on the machine~\cite{Zellweger2016}.
\item  Sampling instructions as they are executed~\cite{Dashti2013} to identify memory loads and stores along with information about where they are located on the system. This can provide very fine grained information about locations of memory accesses.
\item Monitoring page faults to identify suboptimal page placements and configurations~\cite{Lankes2012}.
\item Running benchmarks with known loads and timing their execution to determine latency and or bandwidth properties~\cite{Smccormick2011,Li2013}.  
\end{itemize}
These techniques are applied to many different locations including, improvements to the OS task scheduler, improvements to the OS memory management, and tools and libraries to assist the programmer both by providing information about the expected behavior of the system, and to automate memory placement.

Approaches to scheduling and memory management generally fall into two categories, real time approaches that detect when something is suboptimal, and ahead of time approaches that measure the applications and or system beforehand to provide guidance to the OS. 

Real time approaches such as DINO~\cite{Blagodurov2011} migrate threads and their associated memory pages based on the LLC miss rate, while  Merkel \etal~\cite{Merkel2010} leave memory where it is allocated and just move threads around. Carrefour~\cite{Dashti2013} addresses congestion on memory controllers by sampling the executed instructions observe the loads and stores of different threads. From this it determines if the load on the memory controllers is too high and follows a decision tree to determine if pages should be migrated, interleaved, or replicated to reduce load. Lankes \etal~\cite{Lankes2012} produced suggestions for extending the OS memory management with a page table per NUMA node, allowing pages to be replicated on demand if the load on the interconnect is too high. 

Ahead of time tool such as Pandia~\cite{Goodman:2017:pandia} and libraries like Smart Arrays~\cite{smartArrays}  perform sample runs and provide suggested thread or memory placements,	 these lack  detailed bandwidth models so cannot apply these runs to different thread counts and placementsas effectively.

Many of the approaches require detailed specifications of the system. This is typically generated ahead of the time through the use of benchmarks and either performance counters or wall clock time. Mc~Cormick \etal~\cite{Smccormick2011} measuring the NUMA properties based on  2 benchmarks, measuring latency, and bandwidth to help the scheduler calculate costs and change tasks node/core based on the location of the application memory. Majo~\etal~\cite{Majo2011} took a similar approach using a single memory intensive benchmark.  Pandia uses an extensive set of synthetic benchmarks to measure the memory architecture, as well as the processors compute performance. Li \etal~\cite{Li2013} take a similar extensive approach considering data transfer to disks, networks and GPU's. To calculate the data transfer rates  they copy large amounts of data with \emph{memcpy} and time the transfer. From this they build an NxN matrix. This is expensive to maintain so they categories the machine into groups and provide timings for each group. 

So Many Performance Events, So Little Time~\cite{Zellweger2016} does not model the memory traffic directly, but uses a set of benchmarks which cause specific problems and ML approaches which look at all performance counters. From these they construct a model with a small number of features that can be driven from the counters available to a single  run. The intention of this work is to act as a debugging tool for programmers, with the memory bandwidth and latency issues being detected. As with all ML approaches it requires a sizable training set, and exploring all of the performance counters takes time.

Looking away from the memory interconnects there has been extensive work focusing on the expected behavior of caches and the performance of applications both alone and when sharing a cache. Much of this builds on ideas on measuring the performance of the cache and the applications that are using it, but normally this work constrains itself to just the cache excluding the wider memory system.  Predicting based just on cache interference can be done with mathematical models on data gathered from instrumented workloads~\cite{marin:cross,chandra:predicting,  west:online}. McGregor~\etal~\cite{mcgregor:scheduling}, Knauerhase~\etal~\cite{knauerhase:using}, Fedorova~\etal~\cite{fedorova:improving}, Zhuravlev~\etal~\cite{Zhuravlev2010}, Collins~\etal~\cite{collins:lira} and Dhiman~\etal~\cite{dhiman:vgreen} select threads or workloads to collocate based on performance counters giving measurements such as bus transactions per thread, stall cycles per thread, and LLC miss rate per thread. Xie and Loh classify applications~\cite{xie:animals} based on LLC usage and Lin~\etal~\cite{lin:colors} on L2 usage. These all work to a fixed number of threads, ReSense~\cite{dey:resense} can dynamically controls threads numbers.

\section{Conclusion}
In this paper we have presented and evaluated a model for estimating the bandwidth distribution of analytics applications on NUMA machines. The model is fitted to applications through the use of two instrumented runs with specific thread placements. Testing the results across 1000's of measurements shows a high degree of accuracy for many applications with moderate to high bandwidth requirements, and techniques to detect when the model fails to fit effectively. The ideas presented here have uses ranging from  performance debugging tools, to scheduling tools such as Pandia~\cite{Goodman:2017:pandia}, to libraries that can control memory placements~\cite{smartArrays}.


\end{document}